\documentclass[11pt,a4paper]{article}
\pdfoutput=1
\usepackage{jcappub}
\usepackage{multirow}
\usepackage{lscape}
\usepackage{rotating}
\usepackage{color}
\usepackage{caption}
\usepackage{subcaption}
\usepackage{rotfloat}
\usepackage{upgreek}
\usepackage{wasysym}

\usepackage{aas_macros}

\title{Unidentified sources in the Fermi-LAT second source catalog:
  the case for DM subhalos}

\author{Hannes-S. Zechlin,}
\author{Dieter Horns}
\affiliation{University of Hamburg, Institut f\"ur Experimentalphysik,\\
Luruper Chaussee 149, D-22761 Hamburg, Germany}
\emailAdd{hzechlin@physik.uni-hamburg.de}
\emailAdd{dieter.horns@physik.uni-hamburg.de}

\abstract{The Large Area Telescope (LAT) aboard the \emph{Fermi}
  satellite allows us to study the high-energy $\gamma$-ray sky with
  unprecedented sensitivity. However, the origin of 31\% of the
  detected $\gamma$-ray sources remains unknown. This population of
  unassociated $\gamma$-ray sources may contain new object classes,
  among them sources of photons from self-annihilating or decaying
  non-baryonic dark matter. \emph{Fermi}-LAT might be capable to
  detect up to a few of these dark matter subhalos as faint and
  moderately extended $\gamma$-ray sources with a temporally steady
  high-energy emission. After applying corresponding selection cuts to
  the second year \emph{Fermi} catalog 2FGL, we investigate 13
  candidate objects in more detail including their multi-wavelength
  properties in the radio, infrared, optical, UV, and X-ray bands. For
  the $\gamma$-ray band, we analyze both the 24-month and 42-month
  \emph{Fermi}-LAT data sets. We probe the $\gamma$-ray spectra for
  indications of a spectral cutoff, which singles out four sources of
  particular interest. We find all sources to be compatible with a
  point-source scenario. Multi-wavelength associations and, in
  particular, their infrared color-color data indicate no source to be
  compatible with a dark matter origin, and we find the majority of
  the candidates to probably originate from faint, high-frequency
  peaked BL Lac type objects. We discuss possibilities to further
  investigate source candidates and future prospects to search for
  dark matter subhalos.}


\arxivnumber{1210.3852}

\begin{document}
\maketitle
\flushbottom

\section{Introduction}\label{sec:intro}
Current all-sky surveys of the high-energy (HE) $\gamma$-ray sky
provide unprecedented sensitivity to disseminate the population of
high-energy $\gamma$-ray emitters. Based upon 24 months of data
recorded with the Large Area Telescope (LAT) aboard the \emph{Fermi
  Gamma-ray Space Telescope} satellite (\emph{Fermi})
\cite{2009ApJ...697.1071A}, the \emph{Fermi} Collaboration recently
published the second \emph{Fermi}-LAT point-source catalog (2FGL)
\cite{2012ApJS..199...31N}. The 2FGL contains 1\,873 sources detected
between 100\,MeV and 100\,GeV (with a significance $S \gtrsim
4\sigma$), whereof approximately one third (576 sources) are lacking
reliable association with sources detected in other wavelength
bands. On the contrary, the majority ($\sim$1\,000) of the 1\,297
associated sources have been classified to most likely originate from
active galactic nuclei (AGN), in particular BL Lacs and flat spectrum
radio quasars \cite{2011ApJ...743..171A}.

While it seems plausible that most of the unassociated (high-latitude)
$\gamma$-ray sources are expected to originate from faint AGN, this
sample may also contain new classes of $\gamma$-ray emitting sources
\cite{2010arXiv1007.2644M,2010MNRAS.408..422S,2012ApJ...753...83A,2012ApJ...752...61M,2012arXiv1205.4825M,2012arXiv1209.4359H}. In
particular, this includes sources potentially driven by
self-annihilating (or decaying) dark matter (DM), i.e., DM subhalos
(see \cite{2012A&A...538A..93Z} and references therein). In this
context, it is also interesting to note that the intensity of the
isotropic diffuse $\gamma$-ray background \cite{2010PhRvL.104j1101A}
cannot be fully accounted for by the properties of known $\gamma$-ray
emitters (e.g., blazars) extrapolated below the confusion limit of
\emph{Fermi}-LAT
\cite{2010ApJ...720..435A,2012arXiv1202.5309C}. Unravelling the nature
of \emph{Fermi} unidentified sources therefore remains a crucial task
to tackle searches and constraints on new HE phenomena, e.g.,
self-annihilating DM.

Significant evidence for the existence of a so far undiscovered form
of matter, so-called DM, has been provided by various different
astrophysical observations via its gravitational imprint
\cite{1996PhR...267..195J,2005PhR...405..279B,2010Natur.468..389B}. Cold
DM manifests itself on both cosmological and galactic scales, i.e.,
prevailing the formation of large scale structures down to accounting
for galactic halos and their DM substructure (DM subhalos)
\cite{2008Natur.454..735D,2008Natur.456...73S}. Observations indicate
DM as an unknown non-baryonic type of heavy, electrically neutral, and
color-neutral particle, which is very weakly interacting with standard
model particles. A promising scenario comprises DM to be constituted
by weakly interacting massive particles (WIMPs) of Majorana type, with
a mass between a few hundred GeV and several TeV. With an interaction
strength at the order of weak interactions, thermal production of
WIMPs in the early Universe can consistently explain the measured DM
relic density \cite{2011ApJS..192...18K}. Appropriate WIMP candidates
naturally arise in beyond standard model theories, e.g., Supersymmetry
\cite{1998pesu.conf....1M}.

The discovery of WIMPs is then possible via three complementary
approaches, i.e., collider, direct, and indirect detection techniques:
while WIMPs may be directly produced in colliders with sufficient
center-of-mass energy, e.g., the Large Hadron Collider (LHC)
\cite{2012arXiv1204.5638A,2012arXiv1204.5341C}, underground low-noise
experiments \cite{2012arXiv1203.2566S} are sensitive to their
scattering signatures with heavy nuclei. From the astrophysical point
of view, WIMPs may be indirectly detected through their
self-annihilation (or decay) to standard model final states,
eventually producing $\gamma$ rays, charged light hadrons and leptons,
and neutrinos \cite{2011JCAP...03..051C}.

In this paper, we present a search for DM subhalo candidates in the
2FGL catalog, following up on the 1FGL catalog search we conducted in
\cite{2012A&A...538A..93Z}, henceforth called Paper~I; for related
studies, see
e.g. \cite{2010PhRvD..82f3501B,2011PhRvD..83a5003B,2011arXiv1109.5935N,2012ApJ...747..121A,2012PhRvD..86d3504B,2012arXiv1205.4825M}. The
paper is structured as follows: In section \ref{sec:subhalos}, the
$\gamma$-ray properties of detectable DM subhalos are summarized. The
catalog search for candidate $\gamma$-ray sources and a study of their
spectral and temporal properties are described in section
\ref{sec:2FGL_cand}, including the investigation of multi-wavelength
counterparts and an analysis of archival UV and X-ray data. The
results of this study are summarized and discussed in section
\ref{sec:discussion}.

\section{Gamma-ray emission of DM subhalos}\label{sec:subhalos}
In the hierarchical formation of structures, galactic DM halos are
anticipated to host a large population of smaller subhalos (up to
$10^{16}$). Their mass spectrum $\mathrm{d}N/\mathrm{d}M$ follows a
power-law distribution over the mass range $M$ between
$10^{-11}$--$10^{-3}\,M_\odot$ and $\sim\!10^{10}\,M_\odot$:
$\mathrm{d}N/\mathrm{d}M \propto M^{-\alpha}$, where $\alpha \in
[1.9;2.0]$, see
\cite{2008Natur.454..735D,2008Natur.456...73S,2009NJPh...11j5027B}. While
some of the massive subhalos are expected to host the Milky Way's
luminous dwarf spheroidal satellite galaxies (dSphs), baryonic content
of low-mass subhalos and even concentrated massive subhalos can be
lacking, see \cite{2011MNRAS.415L..40B,2012ApJ...753L..21B} and
references therein. In Paper~I, we have shown that \emph{up to two}
massive DM subhalos between $10^5$ and $10^8\,M_\odot$ could be
detectable with \emph{Fermi}-LAT within the first two mission years,
assuming common models on the self-annihilation of heavy WIMPs, their
density distribution in DM subhalos, and the distribution of the
subhalos in the Galaxy. These objects at distances of
$\mathcal{O}(\mathrm{kpc})$ would appear in the $\gamma$-ray sky as
moderately extended ($\theta_{68} \approx 0.5^\circ$) $\gamma$-ray
sources above 10\,GeV.\footnote{Within the angle $\theta_{68}$ a
  fraction of 68\% of the total $\gamma$-ray luminosity is emitted. In
  comparison with the point spread function of \emph{Fermi}-LAT,
  \mbox{$\theta_{68} = 0.5^\circ$} corresponds to about
  $4\sigma_\mathrm{PSF}$, where $\sigma_\mathrm{PSF} \approx
  0.13^\circ$ at 10\,GeV (see
  http://www.slac.stanford.edu/exp/glast/groups/canda/lat\_Performance.htm).}
The faint, temporally constant $\gamma$-ray flux at energies $E$
between 10 and 100\,GeV is anticipated at the detection level of
\emph{Fermi}-LAT, $\phi_\mathrm{p}(10\!-\!100\,\mathrm{GeV}) \approx
10^{-10}\,\mathrm{cm}^{-2}\,\mathrm{s}^{-1}$, owing to the small
self-annihilation cross section of WIMPs, $\langle \sigma v \rangle
\sim 3\times 10^{-26}\,\mathrm{cm}^3\,\mathrm{s}^{-1}$. The
differential $\gamma$-ray flux approximately follows a hard power-law
(index $\Gamma \lesssim 1.5$) with a distinct cutoff to the WIMP mass
$m_\chi$, where we assumed WIMPs of $m_\chi =
500\,(150)\,\mathrm{GeV}$ annihilating to heavy quarks, gauge bosons
(e.g., $b\overline{b}$, $W^+W^-$), or to the leptons $\tau^+\tau^-$,
see figure 2 in Paper~I. We emphasize that the expected $\gamma$-ray
flux may be even higher when including the possible presence of
sub-substructures (enhancement by a factor of 2 to 3)
\cite{2008ApJ...686..262K,2009JCAP...06..014M} and (rather
model-dependent) photon contributions from final state radiation and
virtual internal Bremsstrahlung
\cite{2005PhRvL..94m1301B,2005PhRvL..95x1301B,2008JHEP...01..049B,2012JCAP...07..054B}.
The rather low WIMP velocity in bound subhalos might also lead to
resonances in the self-annihilation cross section (Sommerfeld
enhancement) \cite{2009PhRvD..79a5014A,2009Sci...325..970K}. Potential
secondary emission from energetic charged leptons eventually produced
by WIMP annihilation in Galactic photon and magnetic fields is
expected to be rather faint and diffuse, e.g.
\cite{2004PhRvD..70b3512B,2007PhRvD..75b3513C,2008ApJ...686.1045J},
and we therefore anticipate DM subhalos to be $\gamma$-ray sources
which are not associated to sources detected in other wavelength bands
at lower energies.

\section{DM subhalo candidates in the 2FGL}\label{sec:2FGL_cand}

\subsection{Source selection}\label{subsec:source_sel}
The selection of candidate objects in the class of unidentified
sources listed in the 2FGL\footnote{Version \emph{v06}} is based upon
the properties of $\gamma$-ray emitting DM subhalos discussed above:
\begin{itemize}
\item[(i)] The sample was reduced to sources at high galactic
  latitudes $|b| \geq 20^\circ$, to avoid confusion with conventional
  Galactic sources and to reduce the impact of diffuse Galactic
  $\gamma$-ray emission.
\item[(ii)] Sources were selected for a steady $\gamma$-ray flux,
  requiring the cataloged variability parameter $var < 41.64$, which
  corresponds to a probability of $P_\mathrm{s} > 1\%$ for the source
  to be steady.
\item[(iii)] To select sources potentially driven by massive WIMPs, a
  detection above 10\,GeV was required. In addition, most of the
  high-energy pulsars located at high Galactic latitude are eliminated
  by this energy cut. Spectrally, $\gamma$-ray pulsars may resemble DM
  subhalos, given their stable emission of characteristically hard
  $\gamma$-ray spectra ($\Gamma < 2$) with typical cutoff energies
  $E_\mathrm{c}$ between $1$ and $10$\,GeV
  \cite{2007ApJ...659L.125B,2010ApJS..187..460A,2012ApJ...748L...2K,2012arXiv1205.3089R}.
\item[(iv)] Spectrally hard sources were selected constraining the
  index of the cataloged power-law fit with $\Gamma < 2.0$.
\end{itemize}
All but the last cut have been already used in Paper~I. Applying cuts
(i) to (iv) to the unassociated sources listed in the 2FGL, 14
unassociated $\gamma$-ray sources remain. With the exception of
2FGL~J2339.6$-$0532, all other sources have a HE flux close to the
detection level of \emph{Fermi}-LAT. This is consistent with the
expectation for candidate sources as discussed in
section~\ref{sec:subhalos}. We therefore discarded the outstandingly
bright object 2FGL~J2339.6$-$0532. Table \ref{tab:candidates} lists
the final sample of 13 candidates sources together with their
positional and spectral properties.
\begin{table}[t]
\begin{center}
 \begin{tabular}{|lcccccc|}
\hline
 \multirow{2}{*}{2FGL name} & $l,b$ & $\sigma_{68}/\sigma_{95}$ & $S$ & \multirow{2}{*}{$\Gamma$} & $\phi_\mathrm{p} (10\!-\!100\,\mathrm{GeV})$ & $S_5$ \\
  & [deg] & [arcmin] & $[\sigma]$ &  & [$10^{-10}\,\mathrm{cm}^{-2}\,\mathrm{s}^{-1}$] & $[\sigma]$ \\
\hline
J0031.0+0724$^{1\mathrm{st}}$ & 114.095,$-$55.108 & 4.4/7.2 & 4.4 & $1.9 \pm 0.3$  & $0.7 \pm 0.3$ & 4.5 \\
J0116.6$-$6153 & 297.749,$-$54.986 & 3.7/6.0 & 5.5 & $1.6 \pm 0.2$  & $0.7 \pm 0.3$ & 4.9 \\
J0143.6$-$5844$^{1\mathrm{st}}$ & 290.468,$-$57.102 & 2.4/3.8 & 14.2 & $1.7 \pm 0.1$  & $2.2 \pm 0.6$ & 9.5 \\
J0305.0$-$1602$^{1\mathrm{st}}$ & 200.151,$-$57.146 & 4.5/7.3 & 5.3 & $1.5 \pm 0.2$  & $0.9 \pm 0.4$ & 4.4 \\
J0312.8+2013 & 162.507,$-$31.569 & 3.7/6.0 & 4.4 & $1.7 \pm 0.2$  & $0.7 \pm 0.3$ & 4.5 \\
J0338.2+1306 & 173.471,$-$32.929 & 3.9/6.3 & 5.8 & $1.5 \pm 0.2$  & $1.1 \pm 0.5$ & 5.1 \\
J0438.0$-$7331 & 286.088,$-$35.168 & 4.1/6.6 & 6.1 & $1.4 \pm 0.2$  & $0.8 \pm 0.4$ & 5.0 \\
J0737.5$-$8246 & 295.086,$-$25.467 & 3.7/6.0 & 4.4 & $1.3 \pm 0.3$  & $1.0 \pm 0.4$ & 5.2 \\
J1223.3+7954 & 124.470,$+$37.134 & 3.6/5.8 & 4.2 & $1.4 \pm 0.3$  & $0.6 \pm 0.3$ & 4.6 \\
J1347.0$-$2956 & 317.047,$+$31.398 & 4.1/6.6 & 5.0 & $1.4 \pm 0.3$  & $1.0 \pm 0.5$ & 4.2 \\
J1410.4+7411 & 115.839,$+$41.825 & 2.9/4.7 & 9.1 & $1.9 \pm 0.1$  & $0.7 \pm 0.3$ & 4.7 \\
J2257.9$-$3646$^{1\mathrm{st}}$ & 3.899,$-$64.186 & 5.0/8.2 & 5.3 & $1.9 \pm 0.2$ & $0.8 \pm 0.4$ & 4.1 \\
J2347.2+0707$^{1\mathrm{st}}$ & 96.214,$-$52.385 & 3.7/5.9 & 7.2 & $2.0 \pm 0.2$ & $0.8 \pm 0.4$ & 4.1 \\
\hline
 \end{tabular}
  \caption{DM subhalo candidates in the 2FGL catalog. The first column
    lists the 2FGL name, where the index ``$1\mathrm{st}$'' flags
    sources which have already been listed in the 1FGL (i.e.,
    1FGL~J0030.7+0724, 1FGL~J0143.9$-$5845, 1FGL~J0305.2$-$1601,
    1FGL~J2257.9$-$3643, and 1FGL~J2347.3+0710). For each source, the
    position is given in galactic coordinates $(l,b)$, together with
    the positional uncertainty $\sigma_{68(95)}$ [68\% (95\%) c.l.,
      semi-major axis], detection significance $S$ in Gaussian sigma,
    the power-law index $\Gamma$, and the photon flux in the
    10--100\,GeV band. The last column $S_5$ lists the significance of
    the 10--100\,GeV detection (in Gaussian
    sigma).} \label{tab:candidates}
 \end{center}
\end{table}

The source 2FGL~J0031.0+0724 has been extensively studied in Paper~I,
where also 2FGL~J0143.6$-$5844 has been listed. An updated discussion
of 2FGL~J0031.0+0724 is presented below. 2FGL~J2257.9$-$3646 has been
claimed as a DM subhalo candidate in \cite{2012PhRvD..86d3504B}, while
very high-energy ($E > 100\,\mathrm{GeV}$) follow-up observations of
2FGL~J2347.2+0707 have been conducted with MAGIC
\cite{2011arXiv1109.5935N}. All remaining sources are new candidates.

\subsection{\emph{Fermi}-LAT data}\label{subsec:LAT_data}
The data analysis of each object in table \ref{tab:candidates} was
based on data recorded with the \emph{Fermi}-LAT in the first 24 as
well as 42 months\footnote{The 42-month data set covers the time
  period between the beginning of August 2008 (239557419 MET) up to
  the beginning of February 2012 (350063020 MET).} of the
mission.\footnote{The LAT data are publicly available at
  http://fermi.gsfc.nasa.gov/ssc/data/.} We chose the same analysis
framework and recommended options that were used for the 2FGL (based
upon 24 months of data), with the exception that the considered energy
range was extended to cover 100\,MeV to 300\,GeV. The data analysis
was performed with the public version of the Fermi Science Tools
(v9r23p1, release date 06 October 2011) using data of Pass-7 event
reconstruction along with the P7\_V6 instrument response
functions.\footnote{See
  http://fermi.gsfc.nasa.gov/ssc/data/analysis/.} All events passing
the SOURCE event class were considered. Events were filtered for a
maximum zenith angle of $100^\circ$ (to eliminate contamination from
the Earth's limb), a maximum rocking-angle of $52^\circ$, and the
recommended quality filters \mbox{DATA\_QUAL == 1} and
\mbox{LAT\_CONFIG == 1} were applied. The (quadratic)
region-of-interest (RoI) was centered on the nominal 2FGL position of
the source-of-interest (SoI) with a size of $20^\circ \times
20^\circ$. The overall spectral fit was performed using the binned
likelihood method (with 10 energy-bins per decade; optimizer
NEWMINUIT, requiring an absolute fit tolerance of $10^{-3}$), where
the source model contained all 2FGL sources within the RoI, along with
fixed cataloged positional and spectral parameters. Including the SoI,
the normalizations $\phi_0$ and indices $\Gamma$ of the default model
of a power-law spectrum [$\mathrm{d}\phi/\mathrm{d}E = \phi_0 \left(
  E/E_0 \right)^{-\Gamma}$] of the innermost six sources were kept
free, while the energy scale $E_0$ was fixed to the cataloged pivot
energy. We used the latest publicly available models for the Galactic
foreground (\emph{gal\_2yearp7v6\_v0.fits}) and isotropic background
emission (\emph{iso\_p7v6source.txt}). The normalization and
corrective power-law index of the Galactic foreground template and the
normalization of the isotropic background template were left free. In
detail, the analysis was performed with the tools \emph{gtselect},
\emph{gtmktime}, \emph{gtbin}, \emph{gtltcube}, \emph{gtexpcube2},
\emph{gtsrcmaps}, and \emph{gtlike}. For each source, we checked that
our overall fit reproduced the cataloged data sufficiently.

The analysis chain for the 42-month data set allowed for a possible
change of the positional coordinates of each SoI. We used
\emph{gtfindsrc} to refit the position. The refined uncertainty
contour (at 95\% confidence level) was computed from the
two-dimensional likelihood function
$\mathcal{L}(\mathrm{RA},\mathrm{Dec})$, requiring $2 \Delta(\log
\mathcal{L} ) = 6.18$ (2 degrees of freedom).

\subsection{Spectral analysis}\label{subsec:spectra}
The energy spectra of the candidate sources have to be consistent with
a spectrum generated by self-annihilating WIMPs. For each candidate
source, we carried out a statistical hypotheses test based upon a
likelihood ratio. The null hypothesis ($H_\mathrm{pl}$) of the SoI to
follow a conventional power-law spectrum was tested against the
hypothesis ($H_\mathrm{exp}$) of a spectrum generated by WIMP
annihilation. The likelihood ratio defines the test statistic
\begin{equation} \label{eq:TSexp}
\mathrm{TS}_\mathrm{exp} = -2 \ln \left(
\frac{\mathcal{L}(H_\mathrm{pl})}{\mathcal{L}(H_\mathrm{exp})} \right),
\end{equation}
where $\mathcal{L}(H)$ denotes the total likelihood for the
corresponding RoI, fitted assuming the SoI to follow the spectral
hypothesis $H$ (cf. section \ref{subsec:LAT_data}).\footnote{Note that
  the spectral models (usually power laws) of all other sources in the
  RoI are kept fixed. Therefore, another possibility would be to use
  the test statistic $\mathrm{TS}$ assigned by the spectral likelihood
  fit (\emph{gtlike}) to measure the source's detection significance,
  $S \propto \sqrt{\mathrm{TS}}$. We checked that both methods give
  similar results, as expected.}

As benchmark models, we probed for WIMP annihilation to heavy quarks
(e.g., $b\overline{b}$), gauge bosons (e.g., $W^+W^-$), and to the
leptons $\tau^+\tau^-$, which lead to a considerably harder
$\gamma$-ray spectrum. For our purpose, the differential $\gamma$-ray
spectra resulting from annihilation of supersymmetric neutralinos to
the afore mentioned final states, see
\cite{2004PhRvD..70j3529F,2011PhRvD..83h3507C}, can be approximated
with a power-law spectrum modified by an exponential cutoff
\cite{1998APh.....9..137B}, $\mathrm{d}N_\gamma/\mathrm{d}x =
N_0\,x^{-\Gamma} \exp \left(-px \right)$, where $x = E/m_\chi$. This
simple parametrization originally introduced for gauge boson final
states \cite{1998APh.....9..137B} also provides a reasonable fit to
the $\tau^+\tau^-$ final states \cite{2011PhRvD..83h3507C} with
different values of $N_0$, $\Gamma$, and $p$. The used parameters are
listed in table~\ref{tab:photon_yields}. This approach simplifies
considerably the treatment and interpretation of the fit
procedure. Given the small number statistics of the faint sources,
more subtle differences in the final state spectrum cannot be
resolved.

\begin{table}[t]
\begin{center}
\begin{tabular}{|lcccc|}
\hline
Channel & $N_0$ & $\Gamma$ & $p$ & Remarks \\
\hline
heavy quarks, & \multirow{2}{*}{0.73} & \multirow{2}{*}{1.5} & \multirow{2}{*}{7.8} & \multirow{2}{*}{\cite{1998APh.....9..137B}} \\
gauge bosons &  &  &  &  \\ 
$\tau^+\tau^-$ & 5.28 & 0.35 & 4.6 & \\
\hline
\end{tabular}
\caption{Fit parameters to approximate the differential photon spectra
  originating from final-state fragmentation of self-annihilating
  neutralinos by power laws with exponential cutoffs. For annihilation
  to heavy quarks and gauge bosons, we use the spectral
  parametrization from \cite{1998APh.....9..137B}, while the photon
  yield from annihilation in $\tau^+\tau^-$ is approximated from
  \cite{2011PhRvD..83h3507C}.}\label{tab:photon_yields}
\end{center}
\end{table}

Therefore, the hypothesis $H_\mathrm{exp}$ was a power-law spectrum
with exponential cutoff, $\mathrm{d}\phi/\mathrm{d}E = \phi_0 \left(
E/E_0 \right)^{-\Gamma} \exp \left( -E/E_\mathrm{c} \right)$. The WIMP
mass is then connected to the cutoff energy via $m_\chi =
p\,E_\mathrm{c}$, and the normalization is $\phi_0 =
N_0\,m_\chi^{\Gamma-1}\,E_0^{-\Gamma}$. The index $\Gamma$ was fixed
to the values given in table \ref{tab:photon_yields}, while the energy
scale was set to $E_0 = 1\,\mathrm{GeV}$. As discussed in section
\ref{subsec:source_sel}, we note that this hypothesis also probes for
$\gamma$-ray pulsars which might contaminate our sample, even though
we expect most of them to be eliminated by the energy cut.

Since $\Gamma$ was kept fixed for $H_\mathrm{exp}$, the null and
alternative hypotheses are not nested. This implies that the test
statistic $\mathrm{TS}_\mathrm{exp}$ does not necessarily follow the
theorems of Wilks \cite{1938Wilks} or Chernoff \cite{1954Chernoff},
i.e., $\mathrm{TS}_\mathrm{exp}$ is not drawn from a chi-square
distribution in the null hypothesis. It also implies that
$\mathrm{TS}_\mathrm{exp}$ can have both negative and positive values:
the hypothesis $H_\mathrm{exp}$ is disfavored if
$\mathrm{TS}_\mathrm{exp} \ll 0$ and favored if
$\mathrm{TS}_\mathrm{exp} \gg 0$. Since the distribution of the test
statistic is a priori not known, the significance of this test had to
be calculated with Monte Carlo simulations. The methods are described
in appendix \ref{app:BMC}. Based upon the simulations, we find that
for the index $\Gamma = 1.5\,(0.35)$, a significance of $2\sigma$
corresponds to $\mathrm{TS}_\mathrm{exp} = -6\,(-20)$ and
$\mathrm{TS}_\mathrm{exp} = 2\,(2)$, respectively, while the $3\sigma$
contour is given by $\mathrm{TS}_\mathrm{exp} = -25\,(-35)$ and
$\mathrm{TS}_\mathrm{exp} = 4\,(7)$; table~\ref{tab:p-value} lists
further significance and their corresponding
$\mathrm{TS}_\mathrm{exp}$ values.

The best-fit parameters for a fixed $\Gamma = 1.5$ and $\Gamma = 0.35$
are summarized in table \ref{tab:probe_spectra_24} for the 24-month
data and in table \ref{tab:probe_spectra_42} for the 42-month
data. Although the fitted spectrum depends on just two free
parameters, the statistical errors of the normalization and cutoff
energy remain comparably large, owing to the sample of very faint
sources studied.\footnote{Note that the errors quoted in tables
  \ref{tab:probe_spectra_24} and \ref{tab:probe_spectra_42} are
  correlated, given that the energy scale $E_0$ was fixed to
  1\,GeV. Choosing the decorrelation energy, in principle, reduces the
  statistical errors.} For the 24-month data, we find that a power-law
with exponential cutoff spectrum is favored for the sources
2FGL~J0305.0$-$1602 and 2FGL~J0338.2+1306, with a significance of
$2.4\sigma$ and $3.3\sigma$, respectively. After 42 months, for both
sources the significance of this initial indication has
decreased. However, an exponential cutoff is now favored for the
sources 2FGL~J0143.6$-$5844 and 2FGL~J1410.4+7411, with significances
of $\sim\!3\sigma$. Vice versa, note that for no source we find such a
spectrum to be disfavored by the data (i.e., by more than $3\sigma$).

The initial indications for an exponential cutoff in the 24-month or
42-month data sets motivate a closer inspection of the sources
2FGL~J0305.0$-$1602, 2FGL~J0338.2+1306, 2FGL~J0143.6$-$5844, and
2FGL~J1410.4+7411. A comprehensive discussion of each source candidate
is given in section \ref{subsec:candidate_disc}.

\begin{table}[t]
\begin{small}
\begin{center}
 \begin{tabular}{|l|ccr|ccr|}
 \hline
  24 months & \multicolumn{3}{c|}{$\Gamma = 1.5$} & \multicolumn{3}{c|}{$\Gamma = 0.35$} \\ \cline{1-1}
 \multirow{2}{*}{2FGL name} & $\phi_0/10^{-11}$ & $E_\mathrm{c}$ & \multirow{2}{*}{$\mathrm{TS}_\mathrm{exp}$} & $\phi_0/10^{-11}$ & $E_\mathrm{c}$ & \multirow{2}{*}{$\mathrm{TS}_\mathrm{exp}$} \\
  & [$(\mathrm{cm}^{2}\,\mathrm{s}\,\mathrm{GeV})^{-1}$] & [GeV] & & [$(\mathrm{cm}^{2}\,\mathrm{s}\,\mathrm{GeV})^{-1}$] & [GeV] & \\
\hline
 J0031.0+0724 & $17.1 \pm 8.0$ & $67 \pm 84$ & $-$2.2 & $1.3 \pm 1.1$ & $21 \pm 12$ & $-$5.8 \\ 
 J0116.6$-$6153 & $20.0 \pm 9.0$ & $34 \pm 33$ & 0.9 & $4.7 \pm 4.2$ & $9 \pm 5$ & $-$0.4 \\
 J0143.6$-$5844 & $70.5 \pm 13.2$ & $46 \pm 25$ & 1.4 & $16.8 \pm 5.5$ & $10 \pm 2$ & $-$17.2 \\
 J0305.0$-$1602 & $21.1 \pm 7.5$ & $70 \pm 78$ & 1.4 & $4.8 \pm 2.7$ & $12 \pm 5$ & 4.1 \\
 J0312.8+2013 & $29.9 \pm 11.7$ & $46 \pm 40$ & $-$0.6 & $2.8 \pm 2.1$ & $16 \pm 8$ & $-$5.2 \\
 J0338.2+1306 & $48.0 \pm 13.5$ & $40 \pm 28$ & 4.2 & $12.6 \pm 5.9$ & $9 \pm 3$ & 6.7 \\
 J0438.0$-$7331 & $24.5 \pm 9.4$ & $105 \pm 134$ & 0.9 & $3.8 \pm 2.5$ & $17 \pm 8$ & 1.1 \\
 J0737.5$-$8246 & $31.6 \pm 11.8$ & $66 \pm 68$ & 1.3& $5.0 \pm 2.8$ & $14 \pm 6$ & 2.6 \\
 J1223.3+7954 & $8.1 \pm 6.0$ & $186 \pm 502$ & 0.1 & $0.3 \pm 0.4$ & $41 \pm 35$ & $-$0.2 \\
 J1347.0$-$2956 & $26.1 \pm 8.6$ & $204 \pm 367$ & 0.2 & $2.0 \pm 1.2$ & $31 \pm 17$ & $-$2.6 \\
 J1410.4+7411 & $40.4 \pm 10.7$ & $30 \pm 18$ & 1.2 & $12.4 \pm 5.3$ & $7 \pm 2$ & $-$3.7 \\
 J2257.9$-$3646 & $33.9 \pm 12.6$ & $17 \pm 12$ & 1.3 & $9.4 \pm 7.0$ & $6 \pm 3$ & $-$4.6 \\
 J2347.2+0707 & $36.6 \pm 10.8$ & $61 \pm 49$ & $-$2.5& $4.7 \pm 2.6$ & $15 \pm 6$ & $-$12.1 \\
\hline
 \end{tabular}
\caption{Best-fit parameters for a power law with exponential cutoff
  spectrum, fitting the 24-month data set between 0.1 and
  300\,GeV. The index was fixed to $\Gamma = 1.5$ or $\Gamma = 0.35$,
  respectively. The table lists the normalization $\phi_0$ and cutoff
  energy $E_\mathrm{c}$. The column $\mathrm{TS}_\mathrm{exp}$ gives
  the likelihood ratio for the comparison with a pure power-law
  fit. See the text for more details on the interpretation of
  $\mathrm{TS}_\mathrm{exp}$.} \label{tab:probe_spectra_24}
 \end{center}
\end{small}
\end{table}
\begin{table}[h]
\begin{small}
\begin{center}
 \begin{tabular}{|l|ccr|ccr|}
 \hline
  42 months & \multicolumn{3}{c|}{$\Gamma = 1.5$} & \multicolumn{3}{c|}{$\Gamma = 0.35$} \\ \cline{1-1}
 \multirow{2}{*}{2FGL name} & $\phi_0/10^{-11}$ & $E_\mathrm{c}$ & \multirow{2}{*}{$\mathrm{TS}_\mathrm{exp}$} & $\phi_0/10^{-11}$ & $E_\mathrm{c}$ & \multirow{2}{*}{$\mathrm{TS}_\mathrm{exp}$} \\
  & [$(\mathrm{cm}^{2}\,\mathrm{s}\,\mathrm{GeV})^{-1}$] & [GeV] & & [$(\mathrm{cm}^{2}\,\mathrm{s}\,\mathrm{GeV})^{-1}$] & [GeV] & \\
\hline
J0031.0+0724 & $17.5 \pm 7.7$ & $38 \pm 35$ & $-$6.9& $1.1 \pm 1.0$ & $18 \pm 11$ & $-$13.9 \\
J0116.6$-$6153 & $27.1 \pm 7.3$ & $44 \pm 30$ & 1.2& $5.8 \pm 3.0$ & $10 \pm 4$ & $-$7.1 \\
J0143.6$-$5844 & $53.6 \pm 8.7$ & $53 \pm 26$ & 3.9& $13.4 \pm 3.9$ & $10 \pm 2$ & $-$17.8 \\
J0305.0$-$1602 &  & $E_\mathrm{c} \rightarrow \infty $ &  & $2.9 \pm 1.7$ & $12 \pm 5$ & 2.3 \\
J0312.8+2013 & $23.4 \pm 8.2$ & $42 \pm 32$ & 0.5& $3.0 \pm 1.8$ & $13 \pm 5$ & $-$3.0 \\
J0338.2+1306 & $32.3 \pm 7.6$ & $153 \pm 177$ & 1.1& $5.8 \pm 2.5$ & $14 \pm 5$ & $-$2.7 \\
J0438.0$-$7331 & $20.5 \pm 7.7$ & $54 \pm 48$ & 0.6& $2.8 \pm 1.9$ & $14 \pm 6$ & $-$3.6 \\
J0737.5$-$8246 & $29.1 \pm 9.4$ & $47 \pm 38$ & 2.2& $6.4 \pm 3.5$ & $10 \pm 4$ & 2.3 \\
J1223.3+7954 & $7.8 \pm 4.9$ & $103 \pm 179$ & $-$0.3& $0.4 \pm 0.4$ & $31 \pm 23$ & $-$2.1 \\
J1347.0$-$2956 & $27.9 \pm 6.8$ & $212 \pm 285$ & 0.3& $2.1 \pm 0.9$ & $31 \pm 13$ & $-$7.5 \\
J1410.4+7411 & $42.0 \pm 8.0$ & $27 \pm 12$ & 4.3& $15.3 \pm 4.7$ & $7 \pm 1$ & $-$3.3 \\
J2257.9$-$3646 & $19.5 \pm 7.9$ & $21 \pm 15$ & 0.8& $3.3 \pm 2.4$ & $9 \pm 4$ & $-$2.8 \\
J2347.2+0707 & $46.9 \pm 9.5$ & $49 \pm 27$ & $-$0.1& $12.0 \pm 4.7$ & $8 \pm 2$ & $-$14.6 \\
\hline
 \end{tabular}
\caption{Best-fit parameters for a power law with exponential cutoff
  spectrum, fitting the 42-month data set between 0.1 and
  300\,GeV. The index was fixed to $\Gamma = 1.5$ or $\Gamma = 0.35$,
  respectively. The table lists the normalization $\phi_0$ and cutoff
  energy $E_\mathrm{c}$. The column $\mathrm{TS}_\mathrm{exp}$ gives
  the likelihood ratio for the comparison with a pure power-law
  fit. See the text for more details on the interpretation of
  $\mathrm{TS}_\mathrm{exp}$.} \label{tab:probe_spectra_42}
 \end{center}
\end{small}
\end{table}

\subsection{Variability and angular extent}\label{subsec:varext}
A $\gamma$-ray signal from a DM subhalo is expected to be constant in
time and may be resolved as an extended source. Therefore, the
temporal and spatial distributions of high-energy photons were tested
for compatibility with a constant flux and the hypothesis of angular
extent. The method used is based upon our previous work in
Paper~I. Most of the source candidates have been detected exclusively
in the upper energy bins with a comparably low background
contamination. Therefore, the tests were applied to the high-energy
photon distribution between 3--300\,GeV and the inclusive interval
10--300\,GeV, respectively. As motivated in section \ref{sec:subhalos}
(for details see Paper~I), high-energy photons within a circular
region of radius $0.5^\circ$ around the nominal 2FGL position were
examined. Due to the low background contamination, this sample is
dominated by signal events (see below).

For the 24-month data, the upper part of table
\ref{tab:HEphotons_2442} lists the number of photons detected from
each source between 3--10\,GeV and 10--300\,GeV, respectively,
together with the energy and event class of the photon with highest
energy. The lower part lists the same quantities for the spectrally
selected list of candidates after 42 months. Additionally, the number
of background photons $N^\mathrm{bg}_\mathrm{pred}$ expected within
$0.5^\circ$ around the source is given for the 10--300\,GeV
interval. These values were derived by fitting a RoI of $1^\circ
\times 1^\circ$ centered on the position of the SoI, fixing the
normalizations (and power-law correction) of the Galactic foreground
and isotropic background templates to the best-fit values obtained
from the entire 0.1--300\,GeV fit.\footnote{Owing to the
  quadratic-shaped RoI, the numbers were multiplied with $\pi\,0.5^2
  \approx 0.785$ to correct for a circular RoI.} On average,
approximately seven photons above 10\,GeV have been detected after 24
months, containing between one to two background photons within a
radius of $0.5^\circ$. We note that only 2FGL~J0338.2+1306 has been
detected above 100\,GeV.

\begin{table}[t]
\begin{center}
 \begin{tabular}{|lccccc|}
 \hline
 \multirow{2}{*}{2FGL name} & \multicolumn{2}{c}{Number of photons ($r \leq 0.5^\circ$)} & $N^\mathrm{bg}_\mathrm{pred}(10\!-\!300\,\mathrm{GeV})$ & $E_\mathrm{max}$ & \multirow{2}{*}{Evcl} \\
  & $3\!-\!10\,\mathrm{GeV}$ & $10\!-\!300\,\mathrm{GeV}$ & gal/iso/$\Sigma$ & [GeV] & \\
 \hline
 J0031.0+0724 &  & 5 & 0.3/0.6/0.9 & $44\pm4$ & 4 \\
 J0116.6$-$6153 &  & 5 & 0.2/0.7/0.9 & $26_{-2}^{+1}$ & 4 \\
 J0143.6$-$5844 & 17 & 15 & 0.2/0.7/0.9 & $45\pm3$ & 4 \\
 J0305.0$-$1602 & 11 & 6 & 0.3/0.6/0.9 & $39_{-1}^{+2}$ & 4 \\
 J0312.8+2013 &  & 5 & 0.9/0.7/1.6 & $35_{-3}^{+4}$ & 4 \\
 J0338.2+1306 & 27 & 8 & 1.3/0.6/1.9 & $29\pm2$ & 4 \\
 J0438.0$-$7331 & 14 & 6 & 0.6/0.8/1.4 & $56\pm3$ & 4 \\
 J0737.5$-$8246 &  & 7 & 0.9/0.6/1.5 & $46_{-4}^{+5}$ & 4 \\
 J1223.3+7954 &  & 6 & 0.7/0.8/1.5 & $61_{-5}^{+6}$ & 4 \\
 J1347.0$-$2956 &  & 6 & 0.7/0.7/1.4 & $55_{-4}^{+6}$ & 2 \\
 J1410.4+7411 & 18 & 6 & 0.4/0.9/1.3 & $35\pm2$ & 4 \\
 J2257.9$-$3646 &  & 6 & 0.2/0.7/0.9 & $18_{-1}^{+2}$ & 4 \\
 J2347.2+0707 & 19 & 5 & 0.7/0.6/1.3 & $84_{-7}^{+10}$ & 4 \\
 \hline
 \multicolumn{1}{c}{} & \multicolumn{1}{c}{} & \multicolumn{1}{c}{} \\
 \hline
 J0143.6$-$5844 & 30 & 23 & 0.5/1.3/1.8 & $53\pm4$ & 4 \\
 J0305.0$-$1602 & 16 & 6 & 0.5/1.0/1.5 & $39_{-1}^{+2}$ & 4 \\
 J0338.2+1306 & 38 & 16 & 2.2/1.0/3.2 & $152_{-13}^{+15}$ & 4 \\
 J1410.4+7411 & 33 & 10 & 0.8/1.6/2.4 & $36\pm2$ & 2 \\
 \hline
 \end{tabular}
 \caption{\textit{Top:} 24-month data: Number of $\gamma$-ray photons,
   listed for detected energy bins between 3--10\,GeV and
   10--300\,GeV, within a radial region of $0.5^\circ$ around the
   source's position. Additionally, the expected numbers of background
   photons from Galactic foreground (gal) and isotropic background
   (iso) between 10--300\,GeV are given. The last columns list the
   energy of the photon with highest energy, and its corresponding
   event classification assigned by LAT data reconstruction (2:
   SOURCE, 4: ULTRACLEAN). \textit{Bottom:} Same as above, for
   42 months of data: Number of $\gamma$-ray photons in the 42-month
   data set for spectrally preselected source candidates, see section
   \ref{subsec:spectra}.} \label{tab:HEphotons_2442}
 \end{center}
\end{table}

The potential variability of the source flux was tested with an
unbinned Kolmogorov-Smirnov (KS) test \cite{2007NR}. This test is
already valid for a small number of photon counts, in distinction to
the binned chi-square method used for the 2FGL catalog. The empirical
cumulative distribution function of the arrival times of individual
photons is compared to a uniform distribution, taking the (possibly
varying) exposure into account. The resulting probabilities
$P_\mathrm{const}$ for the temporal photon distribution to be
consistent with a constant flux are listed in table
\ref{tab:variability} for both the 3--300\,GeV and inclusive
10--300\,GeV interval. In particular, we find 2FGL~J0305.0$-$1602 to
show indications for variability ($P_\mathrm{const}=4\,\permil$).

\begin{table}[t]
 \begin{center}
 \begin{tabular}{|lcccc|}
 \hline
 \multirow{2}{*}{2FGL Name} & \multicolumn{2}{c}{$P_\mathrm{const}$} & \multicolumn{2}{c|}{95\% upper limit $\theta_\mathrm{s}^{95}$ [deg]} \\
  & 3--300\,GeV & 10--300\,GeV & 3--300\,GeV & 10--300\,GeV \\ 
 \hline
 J0031.0+0724 &  & 0.51 &  & 0.53 \\
 J0116.6$-$6153 &  & 0.17 &  & 0.55 \\
 J0143.6$-$5844 & 0.97 & 0.83 & 0.25 & 0.38 \\
 J0305.0$-$1602 & 0.006 & 0.19 & 0.45 & 0.73 \\
 J0312.8+2013 &  & 0.05 &  & 0.50 \\
 J0338.2+1306 & 0.15 & 0.57 & 0.48 & 0.60 \\
 J0438.0$-$7331 & 0.28 & 0.33 & 0.30 & 0.50 \\
 J0737.5$-$8246 &  & 1.00 &  & 0.35 \\
 J1223.3+7954 &  & 0.23 &  & 0.35 \\
 J1347.0$-$2956 &  & 0.40 &  & 0.70 \\
 J1410.4+7411 & 0.78 & 0.66 & 0.38 & 0.60 \\
 J2257.9$-$3646 &  & 0.08 &  & 0.90 \\
 J2347.2+0707 & 0.62 & 0.65 & 0.28 & 0.75 \\
 \hline
 \multicolumn{5}{c}{} \\
 \hline
 J0143.6$-$5844 & 0.89 & 0.46 & 0.17 & 0.25 \\ 
 J0305.0$-$1602 & 0.004 & 0.04 & 0.45 & 0.70 \\
 J0338.2+1306 & 0.02 & 0.82 & 0.23 & 0.38 \\
 J1410.4+7411 & 0.47 & 0.97 & 0.33 & 0.47 \\
 \hline
 \end{tabular}
 \caption{Probability for temporally constant $\gamma$-ray emission
   (left columns) and upper limit on the intrinsic angular extent of
   the signal (95\% c.l., right columns). The quantities are listed
   for the 10--300\,GeV band and, in the case of a detection between
   3--10\,GeV, 3--300\,GeV band. Quantities derived from the
   24\,(42)-month data are shown in the \emph{top} (\emph{bottom})
   panel.}\label{tab:variability}
 \end{center}
\end{table}

We used a likelihood-ratio test to probe for intrinsic spatial
extent. The corresponding likelihood function is given by
\mbox{$L(\theta_\mathrm{s}) = -2 \sum_{i=1}^N \ln \left[
    p_\mathrm{det}({\bf x}_i-\overline{\bf x};\theta_\mathrm{s}) + b
    \right]$}, where $p_\mathrm{det}({\bf x};\theta_\mathrm{s})$
follows the probability distribution function (PDF) for a photon to be
detected at ${\bf x}$, $\overline{\bf x}$ denotes the best-fit
position of the SoI, and $b$ denotes the (flat) PDF of the underlying
background $N_\mathrm{pred}^\mathrm{bg}$
\cite{2012A&A...538A..93Z}. The PDF of ${\bf x}-\overline{\bf x}$ for
an intrinsically extended $\gamma$-ray emitter is the convolution of
\textit{Fermi}-LAT's point spread function (PSF) $p_\mathrm{PSF}$
(version P7\_V6) with the intensity profile $p_\mathrm{int}$ of the
emitter, $p_\mathrm{det} = p_\mathrm{PSF} \ast p_\mathrm{int}$. The
intensity profile $p_\mathrm{int}$ of a DM subhalo follows its
line-of-sight integrated squared density profile. Similar to the
approach followed in Paper~I, the DM density profile of a subhalo was
assumed to follow the spherically symmetric Navarro-Frenk-White (NFW)
profile $\rho(r) \propto \left[ r/r_\mathrm{s}\,(1+r/r_\mathrm{s})^2
  \right]^{-1}$ \cite{1997ApJ...490..493N}, where $r$ denotes the
distance to the center of the halo with a characteristic value at
$r_\mathrm{s}$. Since 87.5\% of the total luminosity are produced
within $r_\mathrm{s}$, it serves as a convenient proxy for the
intrinsic subhalo extent. For a subhalo at distance $D$, this
corresponds to the angle $\theta_\mathrm{s} \approx r_\mathrm{s}/D$.
We remark that 68\% of the total luminosity are produced within
$\theta_{68} \simeq 0.46\,\theta_\mathrm{s}$, which is more convenient
for comparison with observational data.

The minimum $L_\mathrm{min}$ of the likelihood function $L$ arises for
the extension parameter $\theta_\mathrm{s}$ fitting the photon
distribution best. Applying the theorem of Wilks \cite{1938Wilks}, the
quantity $\Delta L = L - L_\mathrm{min}$ follows a chi-square
distribution with one degree of freedom, with additional terms of the
order of $1/N^{1/2}$, which are important for a small number of counts
(see also \cite{1979ApJ...228..939C}).

We find no source candidate which shows indication for an intrinsic
angular extent. Note that this result is in agreement with
ref.~\cite{2012ApJ...747..121A} that also searched for angular
extended sources. For each source, we present upper limits on
$\theta_\mathrm{s}$ at 95\% confidence level in table
\ref{tab:variability}. For the 24-month data between 10--300\,GeV,
note hereby that the given confidence level is not precisely defined,
owing to the low number statistics which might affect the chi-square
distribution. In general, the upper limits range from $0.2^\circ$ to
$0.9^\circ$, where the most constraining ones can obviously be derived
from the largest data sets (42 months, 3--300\,GeV).

\subsection{Multi-wavelength counterparts}\label{ssec:mw_counterparts}
\subsubsection{Catalog data}
Although the sources in table \ref{tab:candidates} are cataloged as
unassociated, we carried out a dedicated counterpart
search.\footnote{Apart from preselected catalogs, the archives
  NASA/IPAC Extragalactic Database (NED, http://ned.ipac.caltech.edu/)
  and HEASARC
  (http://heasarc.gsfc.nasa.gov/cgi-bin/W3Browse/w3browse.pl) were
  queried.} In particular, radio and X-ray sources positionally
located inside the 95\,\%-confidence-level uncertainty contour (listed
in the 2FGL) of the $\gamma$-ray source have been searched for,
providing counterpart candidates in the case of a non-DM origin. The
resulting counterpart candidates are listed in table
\ref{tab:mw_assoc} (see appendix \ref{app:mw_assoc}), ordered by
increasing angular separation from the 2FGL position. For every radio
or X-ray source, the USNO-B1.0 catalog \cite{2003AJ....125..984M} was
searched for a matching optical counterpart.

For all but one source (2FGL~J1410.4+7411), at least one radio source
is located nearby. However, note that owing to the high sensitivity
(therefore low confusion limit) of the NVSS \cite{1998AJ....115.1693C}
and SUMSS \cite{2003MNRAS.342.1117M}, which were used for radio
associations, the high radio source density
($\sim\!0.2\,\mathrm{arcmin}^{-2}$) yields a large number of by-chance
associations in the rather large positional uncertainty of the
$\gamma$-ray sources (see table \ref{tab:candidates} and figure
\ref{fig:skyplots}).

\subsubsection{WISE data} \label{ssec:WISE_assoc}
\begin{figure}[t]
\centering
        \begin{subfigure}[t]{0.48\textwidth}
                \centering
                \includegraphics[width=\textwidth]{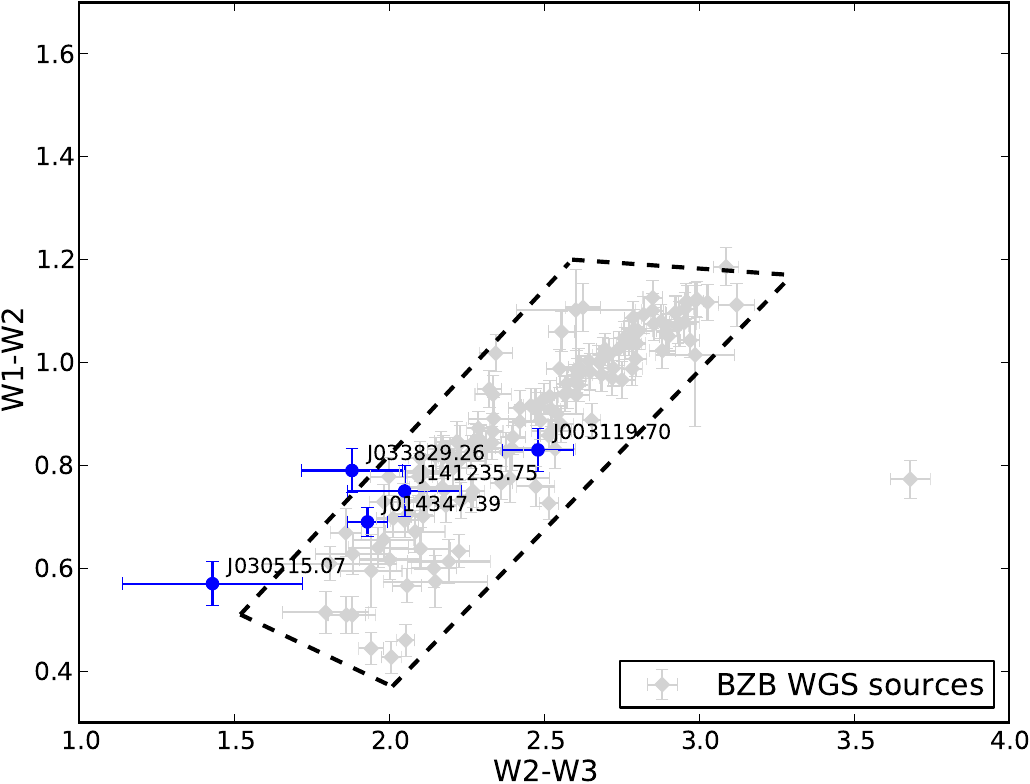}
                \caption{}
                \label{subfig:wise_bzb}
        \end{subfigure}
        \begin{subfigure}[t]{0.48\textwidth}
                \centering
                \includegraphics[width=\textwidth]{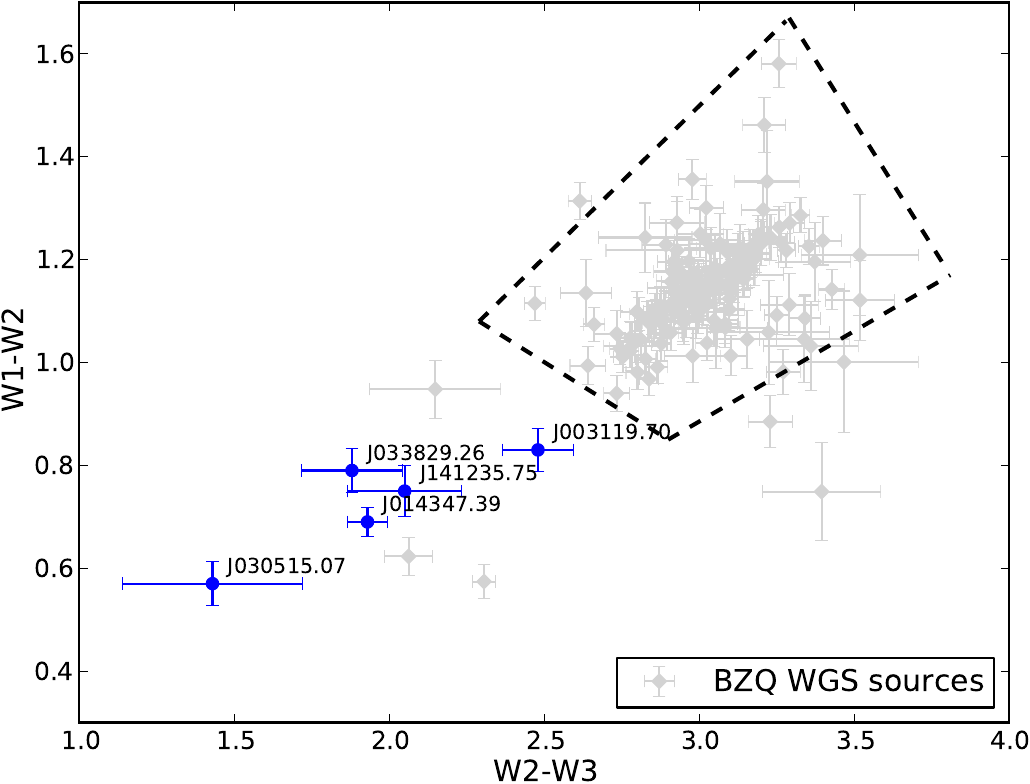}
                \caption{}
                \label{subfig:wise_bzq}
        \end{subfigure}
        \caption{(a): Infrared color-color diagram of the WISE
          associations in table \ref{tab:WISE_assoc} (blue
          circles). The WGS subregion of BZBs is bordered with the
          dashed line, while the reference sources used in
          \cite{2012ApJ...750..138M} are indicated with the gray
          diamonds. (b): Same as (a), comparing to the WGS subregion
          of BZQs.}\label{fig:wise}
\end{figure}

The sources 2FGL~J0312.8+2013, 2FGL~J0737.5$-$8246, and
2FGL~J1347.0$-$2956 have recently been associated with blazar
candidates selected from the WISE survey \cite{2010AJ....140.1868W} on
the basis of their mid-infrared (IR) colors
\cite{2012ApJ...752...61M}. At least one WISE object located in the
positional uncertainty of these $\gamma$-ray sources has been found to
fulfill the IR properties of ($\gamma$-ray) blazars, i.e., lying in
the WISE gamma-ray strip. We adopted this approach
\cite{2012ApJ...752...61M,2012ApJ...748...68D,2012ApJ...750..138M} to
classify IR-counterpart candidates of $\gamma$-ray sources by their
mid-infrared color-color properties. We focus on the IR counterparts
of the four spectrally selected 2FGL candidates, together with the
candidate selected in Paper~I, 2FGL~J0031.0+0724. For each of these
five 2FGL sources, an infrared counterpart candidate is listed in the
WISE catalog \cite{2012yCat.2311....0C}, which is positionally
coincident with the established radio and X-ray associations (see
section \ref{ssec:XRT_ana}, table \ref{tab:mw_assoc}, and section
\ref{subsec:candidate_disc}). The associations are listed in table
\ref{tab:WISE_assoc} (appendix \ref{app:mw_assoc}), together with
their infrared magnitudes $W1, W2, W3, \textnormal{ and } W4$. As
described in \cite{2012ApJ...748...68D,2012ApJ...750..138M} and
references therein, the population of blazars spans a distinct region
in the infrared color space of WISE objects, the WISE Blazar Strip. In
particular, the sample of known $\gamma$-ray emitting blazars
populates the WISE gamma-ray strip (WGS), which is a subspace of the
WISE Blazar Strip. The WGS can be parametrized in two (overlapping)
subregions, one which is dominantly populated by $\gamma$-ray emitting
BL Lacs (BZBs), while the other one is dominated by flat spectrum
radio quasars (BZQs). In figure \ref{fig:wise}, we compare the
infrared color-color diagram ($W2-W3$,$W1-W2$) of the WISE
associations in table \ref{tab:WISE_assoc} with the BZB and BZQ
regions of \cite{2012ApJ...750..138M}.\footnote{Since most of the WISE
  sources tabulated in \ref{tab:WISE_assoc} have not been detected in
  the $W4$ band, we did not attempt to use the color-color projections
  including $W4$.} We find that the WISE associations for all five
2FGL sources hint at a BL Lac origin of the $\gamma$-ray emission, as
being consistent with the BZB region of the WGS.

\subsubsection{\emph{Swift}-UVOT/XRT data} \label{ssec:XRT_ana}
Besides the available catalogs of known X-ray sources, we have
searched for unpublished archival observations of the remaining four
objects. In a dedicated campaign \cite{2011HEAD...12.0403F}, UV and
X-ray follow-up observations of unidentified 2FGL sources have been
carried out with the UVOT and X-ray telescope (XRT,
$0.2\!-\!10\,\mathrm{keV}$) aboard the \emph{Swift} satellite
\cite{2004ApJ...611.1005G,2005ApJ...621..558G}.

The photometric UVOT data were extracted from the products of the
standard pipeline using the HEAsoft~6.12 software package in
combination with the calibration files (2012-04-02). The in-orbit
calibration procedures are described in detail in
\cite{2008MNRAS.383..627P}. The standard aperture of 5 arcsec was used
for all filters to extract the background subtracted flux with
\emph{uvotsource}.

For the corresponding XRT data sets (see table \ref{tab:xrt_data} for
details), calibration and screening (\emph{xrtpipeline}) of the data
acquired in photon-counting (PC) mode was done using standard
screening criteria, along with the current release of calibration
files (2012-04-02). Data were reduced with the HEAsoft~6.11 software
package. We used \emph{Ximage} for source detection and
\emph{Xspec}~(version 12.7.0) for spectral fitting. The probability
limit for a background fluctuation was set to the $5\sigma$-level and
we required a signal-to-noise ratio of $S/N \geq 4$. Positions and
corresponding uncertainties were derived with \emph{xrtcentroid}. For
each source position, ancillary response functions needed for spectral
fitting were derived with \emph{xrtmkarf}, incorporating PSF
correction. The circular on-source regions contained about $90\%$ of
the PSF (corresponding to a radius of $\sim\!47''$), while the
background was derived from appropriate off-source regions with radii
between $3'$ and $5'$. All X-ray sources were spectrally fit with a
power-law model corrected for photoelectric absorption. The hydrogen
column density $N_\mathrm{H}$ was fixed to the nominal Galactic value,
calculated from the Leiden/Argentine/Bonn (LAB) HI survey
\cite{2005A&A...440..775K}. For faint sources ($S/N < 5$), the
power-law index $\Gamma$ was fixed to $2.0$. To achieve sufficient fit
quality, the spectral channels were grouped, requiring a minimum of
$5$ (for $S/N < 15$) or $10$ counts per bin, respectively. We used the
Cash-statistic for spectral fitting, to properly treat the low count
statistic.

\begin{table}[t]
\begin{center}
\begin{tabular}{|lccc|}
\hline
 \multicolumn{1}{|c}{FoV} & Obs. ID & Obs. year & Exposure [ks] \\
\hline
 2FGL J0143.6$-$5844 & 41274 & 2010 & 4.4 \\ 
 2FGL J0305.0$-$1602 & 41286 & 2011 & 3.2 \\
 2FGL J0338.2+1306 & 41292 & 2010 & 4.1 \\
 2FGL J1410.4+7411 & 47219 & 2012 & 3.5 \\
\hline
\end{tabular}
\caption{Archival \emph{Swift}-XRT data for the celestial regions of
  the preselected source candidates. The columns list the observation
  ID, the observation year, and the total exposure in
  ks.}\label{tab:xrt_data}
\end{center}
\end{table}

In general, we find new X-ray sources in every \emph{Swift} field of
view (FoV), with (unabsorbed) fluxes between
$10^{-13}\,\mathrm{erg}\,\mathrm{cm}^{-2}\,\mathrm{s}^{-1}$ and
$10^{-11}\,\mathrm{erg}\,\mathrm{cm}^{-2}\,\mathrm{s}^{-1}$ in the
energy band between 0.3 and 2\,keV. For all selected candidates but
2FGL~J1410.4+7411 some X-ray sources are positionally consistent with
the cataloged $\gamma$-ray uncertainty, and additionally with the
radio detections mentioned above (see table \ref{tab:mw_assoc} in
appendix \ref{app:mw_assoc}). Therefore, they provide convincing
counterpart candidates for the $\gamma$-ray sources, cf. Paper~I. For
reference, positional and spectral parameters of every X-ray detection
are listed in table \ref{tab:swift_assoc} (appendix
\ref{app:mw_assoc}).

\subsection{Discussion of preselected candidates}\label{subsec:candidate_disc} 
Below, we provide a discussion of every preselected source candidate,
based upon the results obtained in sections \ref{subsec:spectra} to
\ref{ssec:mw_counterparts}. Additionally, the updated results on
2FGL~J0031.0+0724 are summarized.

\begin{figure}[t]
\centering
        \begin{subfigure}[t]{0.45\textwidth}
                \centering
                \includegraphics[width=\textwidth]{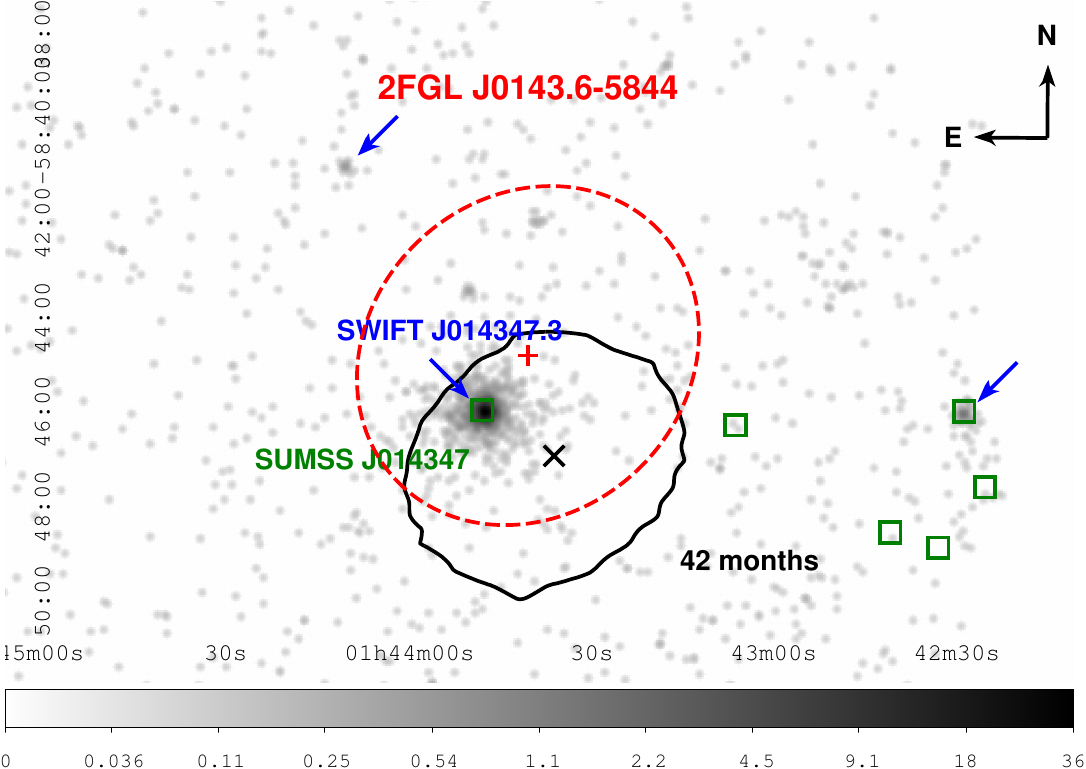}
                \caption{\vspace{0.5cm}}
                \label{subfig:skyplot_J0143.6}
        \end{subfigure}
        \quad\quad\quad
        \begin{subfigure}[t]{0.38\textwidth}
                \centering
                \includegraphics[width=\textwidth]{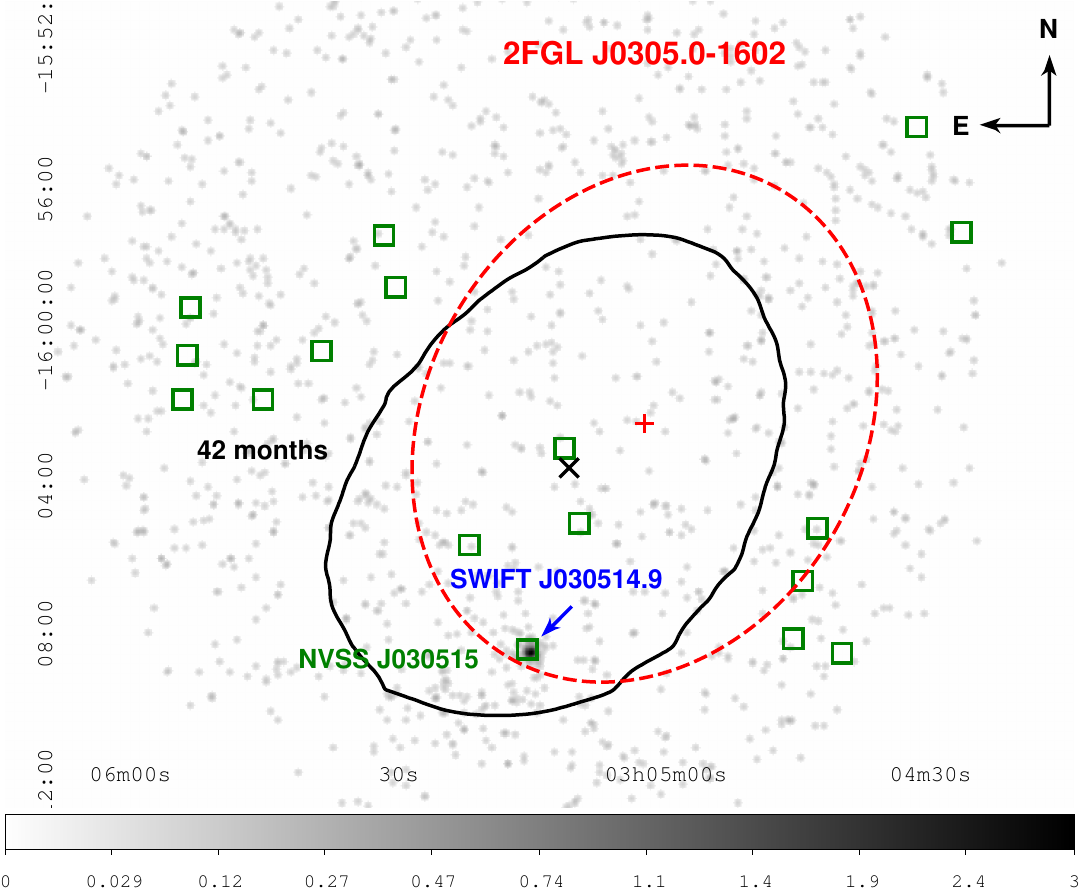}
                \caption{\vspace{0.5cm}}
                \label{subfig:skyplot_J0305.0}
        \end{subfigure}
        \begin{subfigure}[t]{0.38\textwidth}
                \centering
                \includegraphics[width=\textwidth]{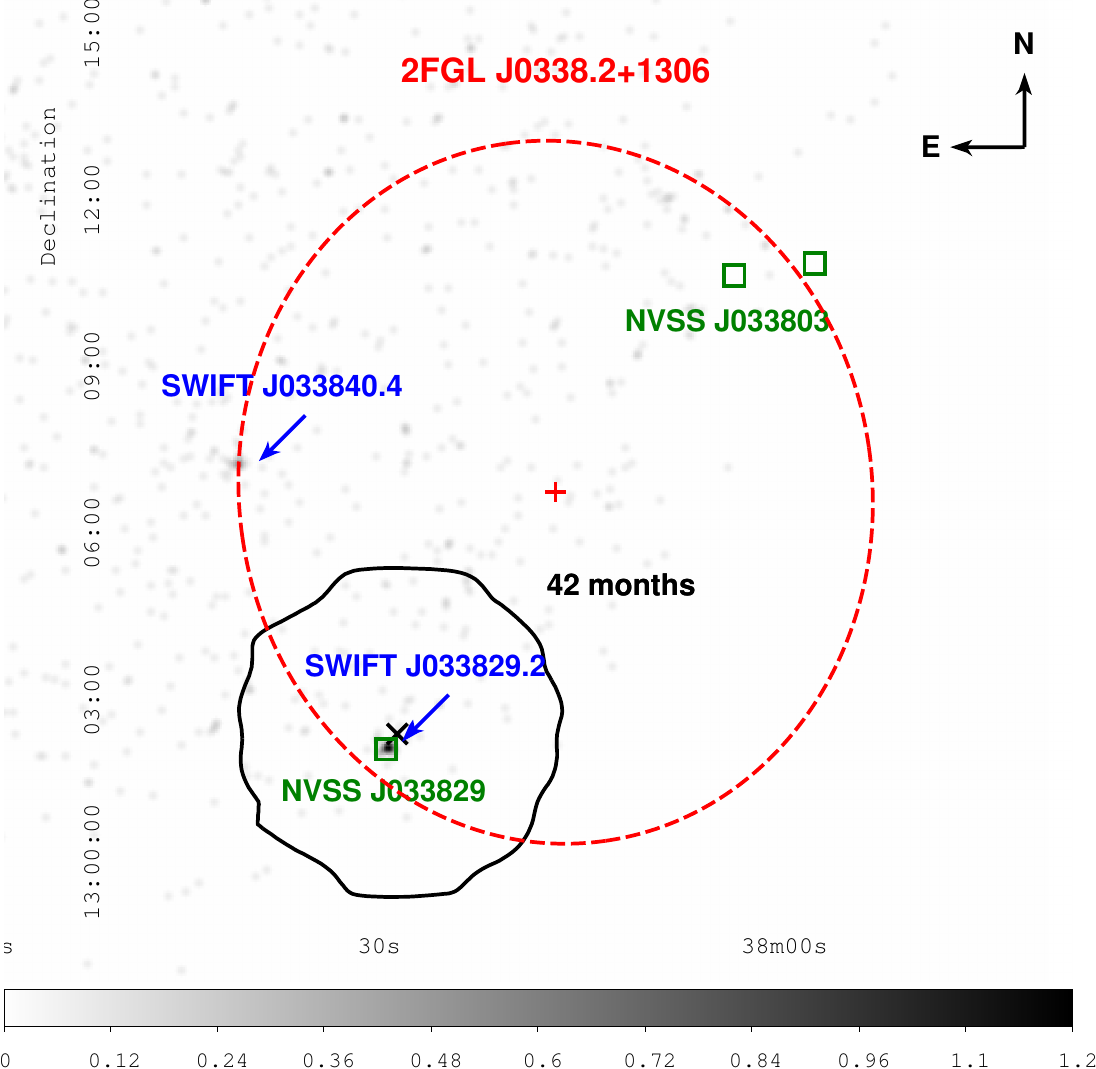}
                \caption{}
                \label{subfig:skyplot_J0338.2}
        \end{subfigure}
        \quad\quad
        \begin{subfigure}[t]{0.5\textwidth}
                \centering
                \includegraphics[width=\textwidth]{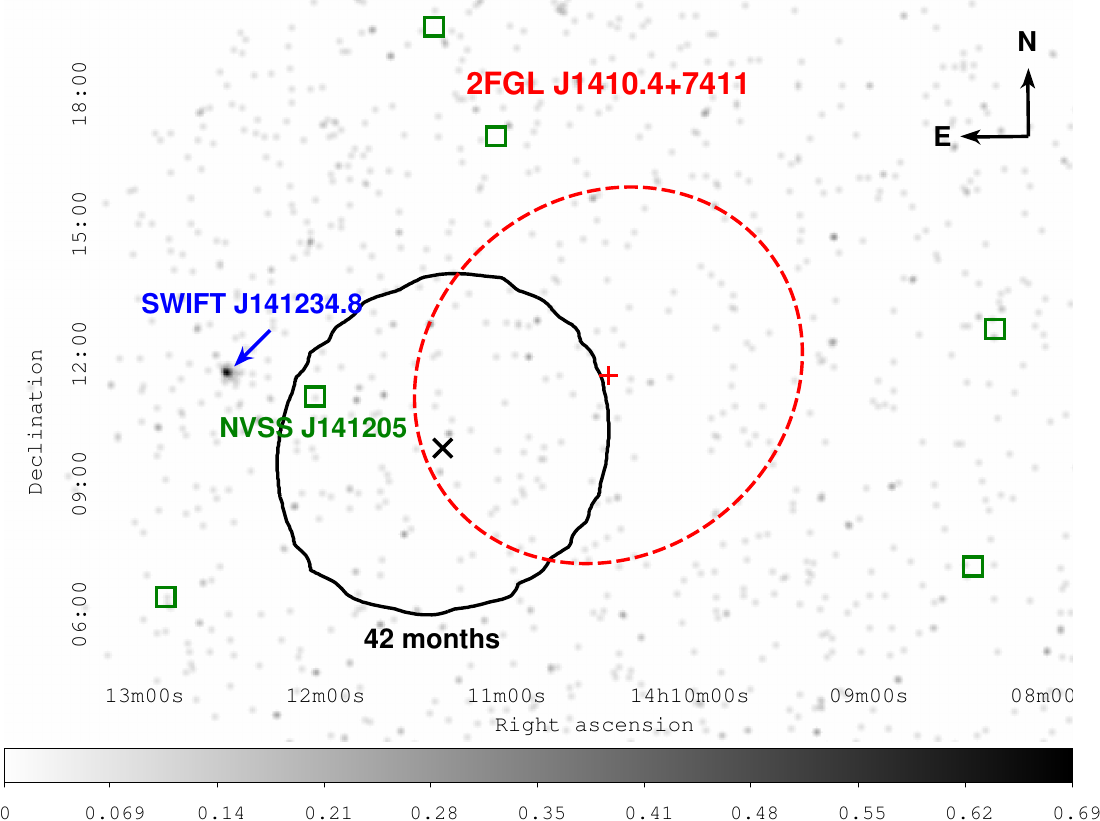}
                \caption{}
                \label{subfig:skyplot_J1410.4}
        \end{subfigure}
        \caption{Best-fit position and uncertainty contour of (a)
          2FGL~J0143.6$-$5844, (b) 2FGL~J0305.0$-$1602, (c)
          2FGL~J0338.2+1306, and (d) 2FGL~J1410.4+7411 for 24 and 42
          months of \emph{Fermi}-LAT data. The cataloged position is
          marked with the red ``$+$'', while the dashed red line
          borders its uncertainty ellipse (95\% c.l.). The black
          ``$\times$'' marks the 42-month position, and the solid
          black line its uncertainty contour (95\% c.l.). An image of
          X-ray photons (\emph{Swift}-XRT), smoothed with a Gaussian
          ($7''$), is shown in the background. Note that the region of
          (c) has not been entirely observed with Swift-XRT, and
          different z-axis scales are used to improve readability
          (i.e., (a) log, (b) sqrt, (c) linear, and (d)
          linear). Positions of radio sources (NVSS) are indicated
          with dark-green boxes, discovered X-ray sources with blue
          arrows. Note that the boxes's size does not reflect the
          positional uncertainty.}\label{fig:skyplots}
\end{figure}

\begin{figure}[t]
\centering
        \begin{subfigure}[t]{0.48\textwidth}
                \centering
                \includegraphics[width=\textwidth]{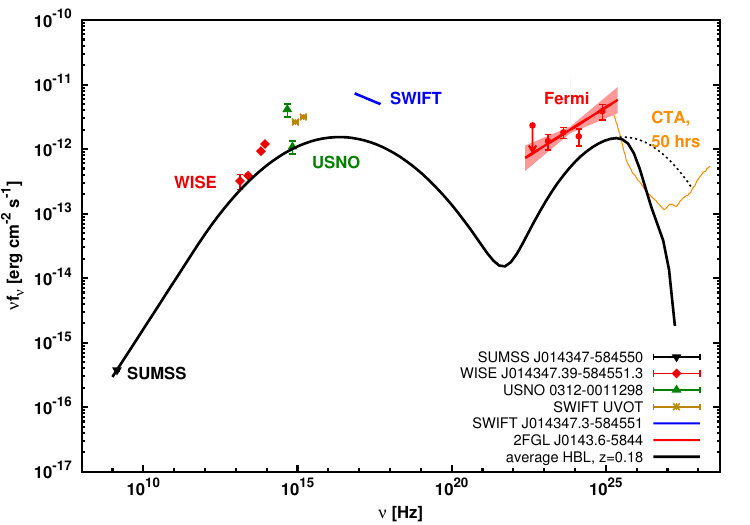}
                \caption{\vspace{0.5cm}}
                \label{subfig:sed_J0143.6}
        \end{subfigure}
        \quad
        \begin{subfigure}[t]{0.48\textwidth}
                \centering
                \includegraphics[width=\textwidth]{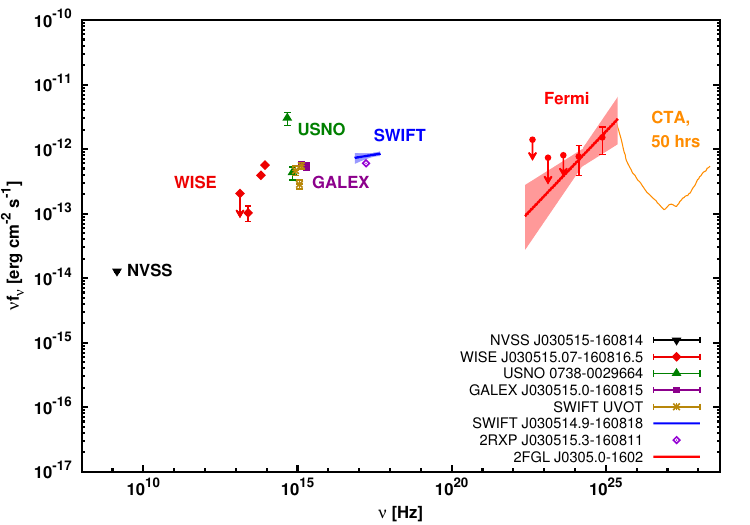}
                \caption{\vspace{0.5cm}}
                \label{subfig:sed_J0305.0}
        \end{subfigure}
        \begin{subfigure}[t]{0.48\textwidth}
                \centering
                \includegraphics[width=\textwidth]{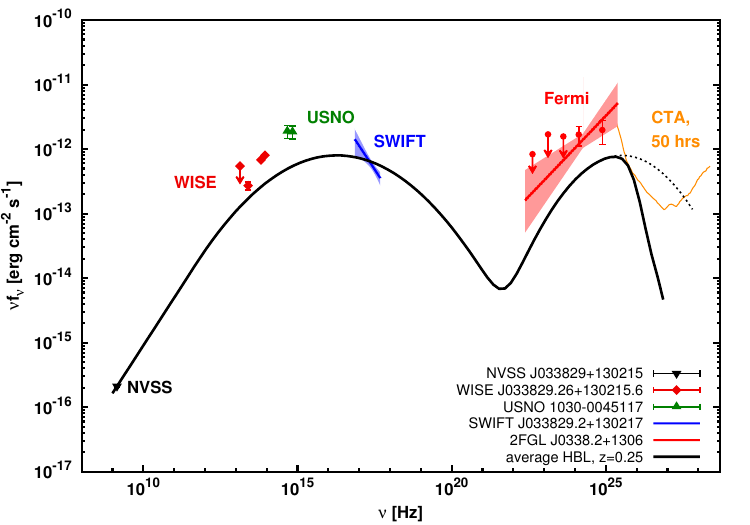}
                \caption{}
                \label{subfig:sed_J0338.2}
        \end{subfigure}
        \quad
        \begin{subfigure}[t]{0.48\textwidth}
                \centering
                \includegraphics[width=\textwidth]{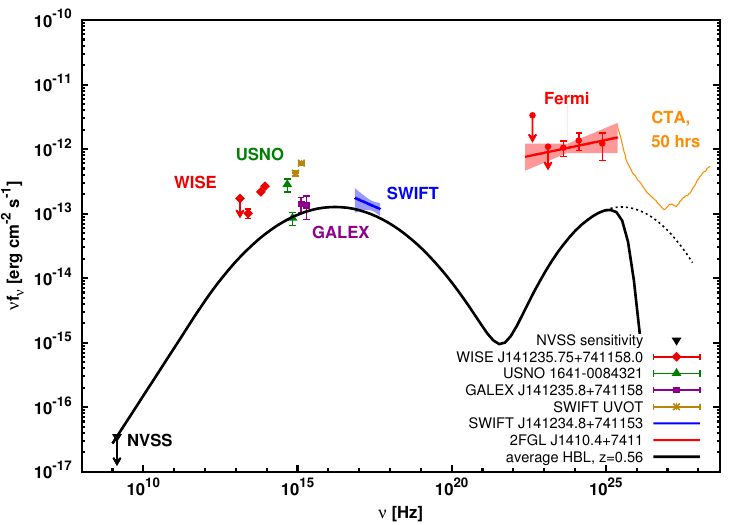}
                \caption{}
                \label{subfig:sed_J1410.4}
        \end{subfigure}
        \caption{Spectral energy distribution (SED) of (a)
          2FGL~J0143.6$-$5844, (b) 2FGL~J0305.0$-$1602, (c)
          2FGL~J0338.2+1306, and (d) 2FGL~J1410.4+7411, assuming the
          multi-wavelength associations discussed in the
          text. Included multi-wavelength data, from low to high
          frequency: radio (NVSS, 1.4\,GHz; black triangle), infrared
          (WISE, $W4,W3,W2,W1$; red diamonds), optical (USNO-B1.0, $R,
          B$; green triangles), ultra-violet (GALEX, NUV, FUV; violet
          boxes; see http://galex.stsci.edu/GR6/; \emph{Swift}-UVOT,
          $U$, $UVW1$, $UVM2$, $UVW2$; dark-golden points), X-ray
          (\emph{Swift}, $0.3\!-\!2\,\mathrm{keV}$; blue line),
          $\gamma$-ray (\emph{Fermi}-LAT 2FGL,
          $0.1\!-\!100\,\mathrm{GeV}$; red line and circles). The
          optical and UV data have been dereddened using $E(B-V)$ from
          \cite{1998ApJ...500..525S} and assuming $R_V=3.1$ (see
          \cite{1989ApJ...345..245C} for details). Arrows indicate
          upper limits (95\% c.l.). Statistical uncertainties of the
          X-ray and $\gamma$-ray spectra are indicated by the
          corresponding shaded areas \cite{2009ApJ...707.1310A}. The
          orange line shows the sensitivity of the planned CTA
          observatory for 50 hours of observation
          \cite{2010arXiv1008.3703C}. For comparison, the solid black
          line shows the average SED of a high-frequency peaked blazar
          (HBL), adapted for the estimated redshifts $z$. The HBL-SED
          is normalized to the radio flux, and the energy flux $\nu
          f_\nu$ is plotted in the frame of a potential observer. The
          HBL-SED has been corrected for EBL absorption (see text for
          details), while the dotted black line shows the SED for a
          vanishing EBL.\vspace{0.5cm}}
        \label{fig:seds}
\end{figure}

Positional and spectral features of the four selected 2FGL candidates
and their corresponding multi-wavelength associations (established in
section \ref{ssec:mw_counterparts}) are summarized in figures
\ref{fig:skyplots} and \ref{fig:seds}, respectively. In figure
\ref{fig:skyplots}, the 2FGL as well as 42-month best-fit position and
their corresponding uncertainties are overlayed over the photon image
measured with \emph{Swift}-XRT, and plotted together with the
positions of radio and X-ray sources. For all four 2FGL sources, the
updated best-fit position shifts by a few arcmins, where the largest
shift was found for 2FGL~J0338.2+1306 ($5.2^\prime$). Figure
\ref{fig:seds} compares the multi-wavelength data with the average
spectral energy distribution (SED) of a high-energy peaked blazar
(HBL), which has been adapted from
\cite{1997MNRAS.289..136F,2001A&A...375..739D} for particular
redshifts $z$. Note in this context that the multi-wavelength data
have not been taken contemporaneously. The redshift of each source was
estimated from the distance modulus $m_\mathrm{R}-M_\mathrm{R}$, where
$m_\mathrm{R}$ denotes the magnitude in the USNO-B1.0 R-band (table
\ref{tab:mw_assoc}), and we assumed the detected optical emission to
originate from a standard giant elliptical host galaxy with an
absolute magnitude $M_\mathrm{R} = -23.1$
\cite{2007A&A...475..199N}. We assumed a vanishing K-correction, i.e.,
a power-law spectrum with index $\alpha = \Gamma - 1 = 1$. We
emphasize that this method only provides a rough estimate under the
given assumptions, while a precise determination of $z$ requires
spectroscopic data in the optical band. In the very high-energy
regime, emitted photons are absorbed through $\gamma$-$\gamma$ pair
production, which was calculated using the EBL model provided in
\cite{2008A&A...487..837F}.

\paragraph{2FGL~J0031.0+0724.}
The 24-month \emph{Fermi}-LAT data of this source have been
intensively studied in Paper~I. The analysis carried out here does not
demonstrate a preference for an exponential cutoff in the 42-month
data set. With a significance of $\sim\!2\sigma$, the lightcurve of
high-energy photons (10--300 GeV) is consistent with a temporally
variable source, and no indication for angularly extended emission was
found. In Paper~I, we already claimed a possible association of the
source with a faint radio source (12\,mJy), positionally coincident
with a faint X-ray source, $f^\mathrm{unabs}(0.2\!-\!2\,\mathrm{keV})
\approx 2.1 \times
10^{-13}\,\mathrm{erg}\,\mathrm{cm}^{-2}\,\mathrm{s}^{-1}$. The
spectral energy distribution (SED) is consistent with a BL Lac origin
(see Paper~I), which is supported by the photometric infrared data of
WISE~J003119.70+072453.6.

\paragraph{2FGL~J0143.6$-$5844.}
The 42-month data indicate the $\gamma$-ray spectrum to be
preferentially fit with an exponential cutoff with
$\sim\!3\sigma$. Furthermore, the lightcurve is preferred to be
steady, but the $\gamma$-ray emission is consistent with a
point-source. Multi-wavelength observations show the source to be
associated with a 27\,mJy radio source (SUMSS), positionally
coincident with the bright X-ray source SWIFT~J014347.3$-$584551,
$f^\mathrm{unabs}(0.3\!-\!2\,\mathrm{keV}) \approx 1.2 \times
10^{-11}\,\mathrm{erg}\,\mathrm{cm}^{-2}\,\mathrm{s}^{-1}$. Together
with the simultaneously measured X-ray flux, the UV data seem to
indicate variable emission when compared with the optical data. The
infrared data support a BL Lac scenario, which is also consistent with
the multi-wavelength SED, see figure \ref{subfig:sed_J0143.6}.

\paragraph{2FGL~J0305.0$-$1602.}
While the $\gamma$-ray data show initial indication (at the
$2.5\sigma$ level) for a spectral cutoff at $\sim\!10\,\mathrm{GeV}$,
the temporal photon distribution of the source excludes a constant
flux with $\sim\!99\%$ confidence. Multi-wavelength searches indicate
the source to be positionally associated with PKS~J0305$-$1608,
showing a radio flux at the Jy level (NVSS) and an X-ray flux of
$f^\mathrm{unabs}(0.3\!-\!2\,\mathrm{keV}) \approx 1.5\times
10^{-12}\,\mathrm{erg}\,\mathrm{cm}^{-2}\,\mathrm{s}^{-1}$. The
photometric infrared data are sparsely consistent with a BL Lac
scenario, while the high radio flux is not in accordance with the
expectation from a high-energy peaked BL Lac (see figure
\ref{subfig:sed_J0305.0}).

\paragraph{2FGL~J0338.2+1306.}
The source was initially selected based upon the 24-month data set,
preferring a spectral cutoff with a significance of
$\sim\!3\sigma$. As shown in figure \ref{subfig:skyplot_J0338.2},
gaining photon statistics revealed a large positional shift fitting
the 42-month data, and the initial indication for a spectral cutoff
vanished. The updated data set also indicates a variable $\gamma$-ray
flux at the $2\sigma$ level. The improved positional accuracy allows
us to associate 2FGL~J0338.2+1306 with a radio source (15.1\,mJy),
which is positionally coincident with SWIFT~J033829.2+130217, see
figure \ref{subfig:skyplot_J0338.2}. Its WISE counterpart suggests a
BL Lac origin, in accordance with the entire multi-wavelength emission
(see figure \ref{subfig:sed_J0338.2}).

\paragraph{2FGL~J1410.4+7411.}
The 42-month $\gamma$-ray data of this source prefer its spectrum to
be fit with a power-law ($\Gamma = 1.5$) with exponential cutoff, with
a significance of $\sim\!3.3\sigma$. The lightcurve is consistent with
steady emission, but no indication for an angular extent was
detected. While also being counterpartless after 24 months, the
position computed from the larger photon sample seems to shift towards
the faint X-ray source SWIFT J141234.8+741153, see figure
\ref{subfig:skyplot_J1410.4},
$f^\mathrm{unabs}(0.3\!-\!2\,\mathrm{keV}) \approx 2.7\times
10^{-13}\,\mathrm{erg}\,\mathrm{cm}^{-2}\,\mathrm{s}^{-1}$.\footnote{Note
  that the X-ray source XMMSL1~J141002.6+740744
  ($f^\mathrm{unabs}(0.2\!-\!2\,\mathrm{keV}) \approx 2 \times
  10^{-12}\,\mathrm{erg}\,\mathrm{cm}^{-2}\,\mathrm{s}^{-1}$)
  \cite{2008A&A...480..611S} is located just outside the 2FGL $95\%$
  uncertainty contour, but does not appear in the SWIFT observations.}
Note that the \emph{Swift} source has no radio counterpart, which
might reflect the lacking sensitivity of radio surveys. Assuming the
\emph{Swift} source to be the correct X-ray association, its infrared
counterpart WISE J141235.75+741158.0 would indicate a BL Lac
origin. Additionally, compared to the USNO and GALEX data, the
\emph{Swift}-UVOT observations indicate variable emission.

\section{Discussion and conclusions}\label{sec:discussion}
In this work, we investigated the unassociated $\gamma$-ray source
population of the \emph{Fermi}-LAT second year point-source catalog
for sources potentially originating from DM subhalos. Basic catalog
selection revealed 13 high-latitude sources, with hard $\gamma$-ray
spectra detected above 10\,GeV, and lacking indication for temporal
variability. Using 3.5~years (42~months) of \emph{Fermi}-LAT data, we
developed a statistical test to probe the candidates for spectral
consistency with self-annihilating DM (a power-law with exponential
cutoff). The high-energy spectra of a subset of 4 sources were found
to be preferentially fit by power-laws with exponential cutoff (i.e.,
2FGL~J0143.6$-$5844, 2FGL~J0305.0$-$1602, 2FGL~J0338.2+1306, and
2FGL~J1410.4+7411), with significances between $2.5\sigma$ and
$3.3\sigma$. All sources were tested for temporally constant and
spatially extended $\gamma$-ray emission. The $\gamma$-ray emission of
2FGL~J0305.0$-$1602 shows a $\sim\!3\sigma$ indication to be
temporally variable, while none of the 13 sources shows indications
for angularly extended emission. Multi-wavelength studies were
conducted to search for associations, using refined positional
information based upon 3.5 years of \emph{Fermi}-LAT data. For
2FGL~J0143.6$-$5844, 2FGL~J0305.0$-$1602, and 2FGL~J0338.2+1306, we
established clear associations detected in the radio, infrared,
optical, UV, and X-ray bands, while 2FGL~J1410.4+7411 is indicated to
be associated to a faint X-ray source, which has also been detected in
the infrared, optical, and UV band.

With the exception of 2FGL~J0305.0$-$1602, the infrared color-color
data of the three other sources is fully consistent with the
population of BL Lacs detected with \emph{Fermi}-LAT. In addition,
such a scenario would be favored by the multi-wavelength data, in
particular by the faint radio, X-ray, and hard $\gamma$-ray emission,
indicating a scenario of a high-frequency peaked BL Lac (see figure
\ref{fig:seds} and Paper~I for details). For all three cases, we note
that the $\gamma$-ray flux predicted by the average SED is below the
\emph{Fermi}-LAT measurement. This might indicate a sample bias,
meaning that the sources have been detected in a flaring state. Within
the errors, the infrared association of 2FGL~J0305.0$-$1602 might also
indicate a BL Lac origin, while in particular its bright radio
counterpart may point towards a different scenario. Finally, the
recent study in \cite{2012arXiv1205.4825M} has attempted to classify
the entire sample of unassociated \emph{Fermi}-LAT sources,
distinguishing between AGN-like and pulsar-like sources. Using a
Random Forest (RF) classifier trained on cataloged $\gamma$-ray
properties of associated \emph{Fermi}-LAT sources, the major fraction
of the unassociated sources are predicted to be AGN, and no
significant outliers have been found. All 13 candidates we selected in
table \ref{tab:candidates} are suggested to be AGN. Consistently, the
recent RF classification in \cite{2012arXiv1209.4359H} assigns all of
them to originate from BL Lacs. In particular, this strongly supports
the BL Lac origin of the four spectrally selected candidates. Note,
however, that since we predict $\gamma$-ray sources originating from
DM subhalos to be particularly faint, their cataloged spectral and
localization parameters suffer from large statistical
uncertainties. Therefore, DM subhalos might hide in the sample of
sources classified by the RF algorithm, emphasizing the necessity of
the in-depth investigations carried out in this paper.

In conclusion, we find no unassociated $\gamma$-ray source in the 2FGL
catalog which is favored to originate from a subhalo driven by
self-annihilating WIMPs at a mass scale above 100\,GeV. However, we
conclude that, among all candidates, the source 2FGL~J1410.4+7411
would be the most interesting, owing to $\gamma$-ray properties which
might prefer a DM origin and a high uncertainty about its
association. From the final source sample we can exclude
2FGL~J0305.0$-$1602 (owing to variable $\gamma$-ray emission). We note
that no source can be firmly excluded by spectral properties. We find
the remaining candidates to most likely originate from faint BL Lacs.

Albeit our prediction of a BL Lac origin, we investigated all source
candidates in the context of the recently claimed evidence for a
line-like feature at $\sim\!130\,\mathrm{GeV}$ in the Galactic Center
region
\cite{2012JCAP...07..054B,2012JCAP...08..007W,2012arXiv1206.1616S,2012arXiv1209.4562F,2012arXiv1209.4548H,2012arXiv1207.6773B,2012JCAP...09..032T,2012arXiv1207.4466H}. In
the case of a self-annihilating DM scenario, DM subhalos will also
appear with a $\gamma$-ray line at $\sim\!130\,\mathrm{GeV}$, and
searches have been started in
\cite{2012arXiv1207.7060S,2012arXiv1208.0828H,2012arXiv1208.1693M,2012arXiv1209.4548H}. Except
for 2FGL~J0338.2+1306, having a photon at $152\,\mathrm{GeV}$, we note
that none of the other sources has been detected above
$100\,\mathrm{GeV}$.

The study presented here has clearly outlined the problems of
identifying faint \emph{Fermi} sources. Difficulties are mainly
related to the limiting collection area of the LAT at the high energy
end, resulting in a small number of photons. This implies the
consequently large uncertainty of source locations, which in turn
leads to source confusion at the faint end of source
populations. Likewise, the small number of photons limits the ability
to determine spectral and temporal properties at the high energy end
of the \emph{Fermi}-LAT response with sufficient accuracy to
distinguish source models. Finally, from the theoretical point of view
in the considered DM scenario, the entirely limited number of
detectable subhalos prohibits conclusive population studies.

Focussing on high-latitude sources with fluxes at the level of the
studied ones, we emphasize that at least some of these issues can be
addressed with larger data sets based upon longer observations. The
correspondingly larger signal-to-background ratios allow us to improve
the positional accuracy by a factor of $\sim$2 with 10~years of
\emph{Fermi}-LAT data, for instance, and to monitor the temporal
photon distribution over longer time periods. The improvement in
sensitivity might push the number of detectable subhalos to
$\mathcal{O}(5)$.

In particular, the issues can also be addressed with pointed follow-up
observations in the very high-energy (VHE) band. The large collection
areas provided by imaging air Cherenkov telescopes (IACTs) for
energies above $\sim\!50\,\mathrm{GeV}$, such as H.E.S.S.-II
\cite{2005ICRC....5..163V,2006A&A...457..899A}, MAGIC stereo
\cite{2008ApJ...674.1037A,2010NIMPA.623..437F}, VERITAS
\cite{2002APh....17..221W,2011arXiv1111.1225H}, and, in particular,
the planned Cherenkov telescope array (CTA)
\cite{2010arXiv1008.3703C,2011NIMPA.630..285D,2012arXiv1208.5356D},
allow the detection of a larger photon sample, significantly reducing
positional and spectral uncertainties. If detected in the VHE, source
candidates therefore can be easily associated or even identified, see
\cite{2011arXiv1109.5935N}.

As a final remark, a potential successor of \emph{Fermi}-LAT such as
GAMMA-400 \cite{2012arXiv1201.2490G,2012arXiv1210.1457G} will
significantly improve the observable energy range
($100\,\mathrm{MeV}\!-\!3\,\mathrm{TeV}$), angular resolution
($\sim\!0.01^\circ$ at $100\,\mathrm{GeV}$), and energy resolution
($\sim\!1\%$ at $100\,\mathrm{GeV}$). The launch of GAMMA-400 is
planned for 2018. For the case of unidentified \emph{Fermi}-LAT
sources, such a telescope will constrain their celestial position with
enhanced precision.

\acknowledgments HSZ kindly acknowledges Francesco Massaro for
providing data published in \cite{2012ApJ...750..138M}, Jules\,P.
Halpern, and Christoph Weniger for helpful discussions. We acknowledge
the anonymous referee for useful comments. This work was supported
through the collaborative research center (SFB) 676 ``Particles,
Strings, and the Early Universe'' at the University of Hamburg. This
research has made use of the NASA/IPAC Extragalactic Database (NED)
which is operated by the Jet Propulsion Laboratory, California
Institute of Technology, under contract with the National Aeronautics
and Space Administration.

\begin{appendix}
\section{Probability distribution of $\mathrm{TS}_\mathrm{exp}$} \label{app:BMC}
To determine the probability density distribution (pdf) of
$\mathrm{TS}_\mathrm{exp}$ (eq. \ref{eq:TSexp}) in the null-hypothesis
$H_0$, we used bootstrap Monte Carlo simulations
\cite{2007NR}. Assuming a pure power-law spectrum (i.e., $H_0$), we
simulated data of the RoI corresponding to 2FGL~J0338.2+1306, which
was exemplarily selected to check whether an exponential cutoff is
preferred by the actual data set (see table
\ref{tab:probe_spectra_24}). The $25\,000$ simulated data sets were
then analyzed with the framework described in section
\ref{subsec:LAT_data} and \ref{subsec:spectra}, therefore calculating
$\mathrm{TS}_\mathrm{exp}$ in exactly the same way as in the actual
data analysis. Each individual analysis procedure was checked for
non-converging behavior ($\sim\!14\%$), and the final data set of
simulations was cleaned accordingly.

The simulations were done in a two-step approach: First, we used
\emph{gtobssim} to simulate five individual 24-month data sets of the
RoI centered on 2FGL~J0338.2+1306 (between 100\,MeV and 300\,GeV). All
sources in the RoI were modeled with the cataloged power-law spectra,
since the current version of \emph{gtobssim} does not accept
log-parabola spectra (which are sometimes preferred). Background
models were implemented as described in section \ref{subsec:LAT_data},
and we used the actual spacecraft file of the first 24 months of
data-taking. We checked the consistency between the analysis results
from the simulated data and the actual data set. As a second step, the
five individual simulations were merged to one data set. Applying the
bootstrap technique to the merged data, we generated $25\,000$
individual 24-month data sets to be used in our analysis. Again, we
checked that the analysis reproduced the actual data well.

\begin{figure}
\centering
        \begin{subfigure}[t]{0.35\textwidth}
                \centering
                \includegraphics[width=\textwidth]{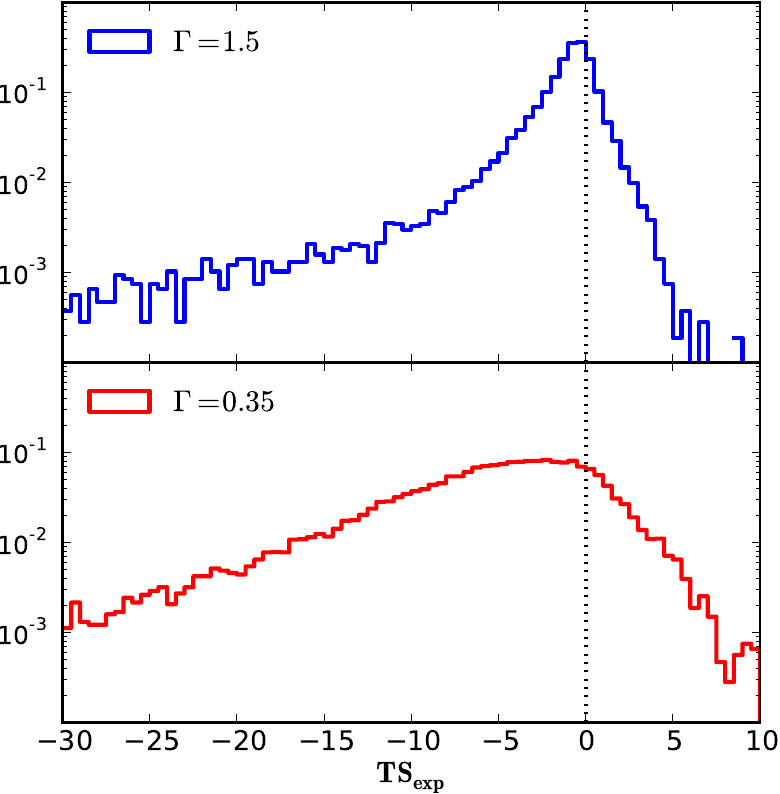}
                \caption{}
                \label{subfig:TSexp}
        \end{subfigure}
        \quad\quad
        \begin{subfigure}[t]{0.47\textwidth}
                \centering
                \includegraphics[width=\textwidth]{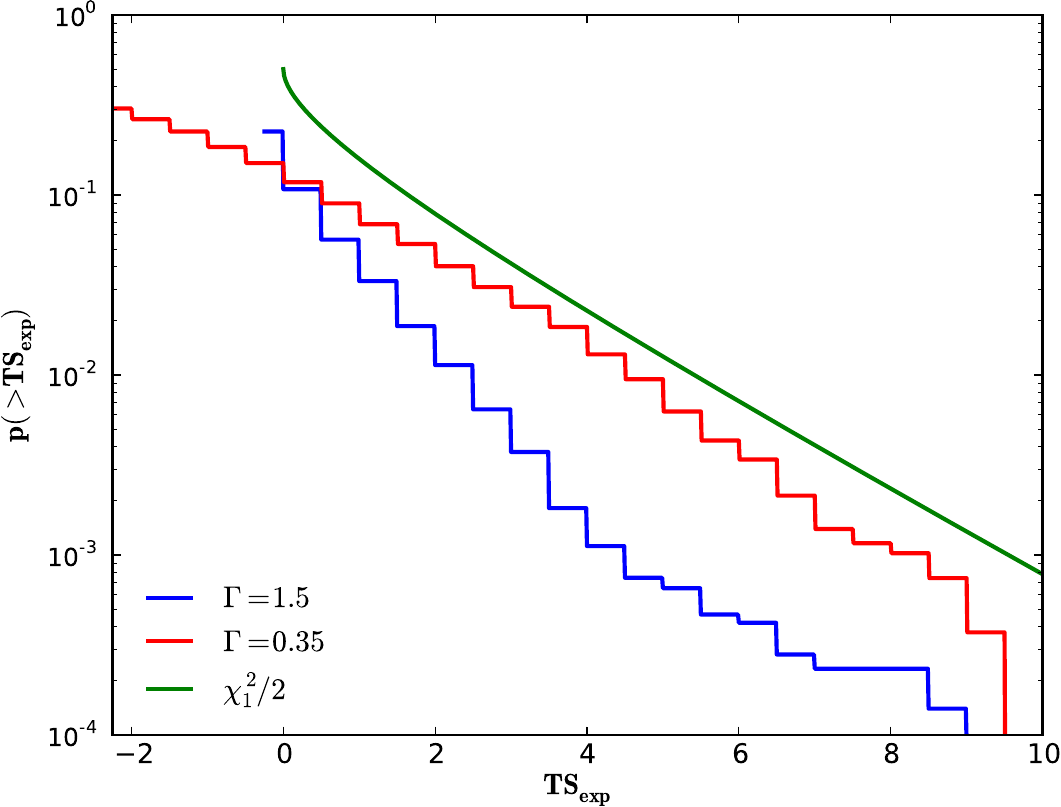}
                \caption{}
                \label{subfig:p-value}
        \end{subfigure}
        \caption{(a): Probability density distribution of the test
          statistic $\mathrm{TS}_\mathrm{exp}$ in the null hypothesis
          $H_0$. The power-law index of the alternative hypothesis
          $H_1$ was fixed to $\Gamma = 1.5$ (\emph{top}) and $\Gamma =
          0.35$ (\emph{bottom}). The dotted vertical line indicates
          $\mathrm{TS}_\mathrm{exp} = 0$. (b): P-value
          $p(>\mathrm{TS}_\mathrm{exp})$ for the positive half-plane
          of the distribution, starting from its corresponding
          maximum. The curves for $\Gamma = 1.5$ (blue line) and
          $\Gamma = 0.35$ (red line) are compared to a $\chi_1^2/2$
          distribution (green line).}\label{fig:TSexp}
\end{figure}

Figure \ref{subfig:TSexp} shows the pdf of $\mathrm{TS}_\mathrm{exp}$,
fixing the index $\Gamma$ of the alternative hypothesis $H_1$
(power-law with exponential cutoff) to $\Gamma=1.5$ and $\Gamma=0.35$,
respectively. The simulation shows that the pdf does indeed not follow
a $\chi^2/2$-distribution \cite{1954Chernoff} for both negative and
positive $\mathrm{TS}_\mathrm{exp}$ values. Rather, we find asymmetric
pdfs with maxima and large tails at negative
$\mathrm{TS}_\mathrm{exp}$ values.\footnote{The maxima are at
  $\mathrm{TS}_\mathrm{exp} = -0.25\,(-2.25)$ for $\Gamma =
  1.5\,(0.35)$.} These features are more pronounced for $\Gamma =
0.35$, owing to the sharply peaked $\gamma$-ray spectrum. For the
positive half-plane, the $p$-values $p(>\mathrm{TS}_\mathrm{exp})$
(the significance of the test statistic) are shown in figure
\ref{subfig:p-value}. The $p$-values are compared to the prediction of
Chernoff's theorem \cite{1954Chernoff}, where the test statistic in
the positive half-plane should follow $0.5\left[
  \delta(\mathrm{TS}_\mathrm{exp}) +
  \chi_1^2(\mathrm{TS}_\mathrm{exp}) \right]$, with $\delta(x)$ the
delta-function and $\chi_1^2(x)$ the chi-square distribution with one
degree of freedom. For $\Gamma = 0.35$, we find that for large,
positive $\mathrm{TS}_\mathrm{exp}$ the pdf approximately follows
Chernoff's theorem, while the pdf of a $\Gamma = 1.5$ spectrum does
not. The resulting significances corresponding to selected
$\mathrm{TS}_\mathrm{exp}$ values are shown in table
\ref{tab:p-value}. For the index $\Gamma = 1.5\,(0.35)$, a
significance of $2\sigma$ corresponds to $\mathrm{TS}_\mathrm{exp} =
-6\,(-20)$ and $\mathrm{TS}_\mathrm{exp} = 2\,(2)$, respectively,
while the $3\sigma$ contour is given by $\mathrm{TS}_\mathrm{exp} =
-25\,(-35)$ and $\mathrm{TS}_\mathrm{exp} = 4\,(7)$.

\newpage
\begin{table}[t]
\centering
\begin{tabular}{|lcc||lcc|}
\hline
\multicolumn{3}{|c||}{$p(<\mathrm{TS}_\mathrm{exp})\,[\sigma]$} & \multicolumn{3}{c|}{$p(>\mathrm{TS}_\mathrm{exp})\,[\sigma]$} \\
$\mathrm{TS}_\mathrm{exp}$ & $\Gamma = 1.5$ & $\Gamma = 0.35$ & $\mathrm{TS}_\mathrm{exp}$ & $\Gamma = 1.5$ & $\Gamma = 0.35$ \\
\hline
$-$40.0 &  & 3.9 & 0.0 & 1.3 & 1.5 \\
$-$35.0 &  & 3.0 & 2.0 & 2.4 & 2.0 \\
$-$30.0 & 3.8 & 2.7 & 4.0 & 3.2 & 2.4 \\
$-$25.0 & 3.0 & 2.4 & 6.0 & 3.5 & 2.9 \\
$-$20.0 & 2.7 & 2.1 & 8.0 & 3.7 & 3.3 \\
$-$15.0 & 2.5 & 1.8 & 10.0 & 4.0 & 4.1 \\
$-$10.0 & 2.3 & 1.3 &  &  &  \\
$-$5.0 & 1.8 & 0.7 &  &  &  \\
0.0 & 0.2 & 0.2 &  &  &  \\
\hline
\end{tabular}
\caption{Probability ($p$-value) to randomly find the test statistic
  to be lower (\emph{left table}) or larger (\emph{right table}) than a
  certain $\mathrm{TS}_\mathrm{exp}$ value, given in Gaussian
  sigma. The $p$-value is listed for both assuming the index $\Gamma$
  of the power law with exponential cutoff to be 1.5 and 0.35,
  respectively.} \label{tab:p-value}
\end{table}

\section{Multi-wavelength association} \label{app:mw_assoc}
\textheight=21.5cm
\begin{sidewaystable}[H]
\begin{scriptsize}
\begin{center}
 \begin{tabular}{|lllllc|}
\hline
 2FGL name & Sep. & Radio & Optical (USNO-B1.0) & X-ray & WGS \\
\hline
\multirow{4}{*}{J0031.0+0724} & 
 $3.1^\prime$ & -- & 0973-0005560: $19.5^\mathrm{m}/18.2^\mathrm{m}$ & SWIFT J003054.9+072328: $0.03(1)$ & \\
 & $3.4^\prime$ & NVSS J003119+072456: $12(1)\,\mathrm{mJy}$ & 0974-0005617: $19.8^\mathrm{m}/18.6^\mathrm{m}$ & SWIFT J003119.8+072454: $0.21(7)$ & \boldmath$\surd$ \\
 & $4.5^\prime$ & -- & [HB2010a] J007.70635+07.38744: $20.9^{\mathrm{m}\,\star}$ & SWIFT J003049.8+072316: $0.04(1)$ & \\
 & $6.2^\prime$ & NVSS J003128+072204: $12(1)\,\mathrm{mJy}$ & -- & -- & \\
\hline
\multirow{5}{*}{J0116.6$-$6153} &
 \multirow{2}{*}{$2.7^\prime$} & \multirow{2}{*}{SUMSS J011619$-$615343: $24(1)\,\mathrm{mJy/beam}^\S$} & 0281-0014602: $18.2^\mathrm{m}/17.8^\mathrm{m}$ & \multirow{2}{*}{--} & \\
  &  &  & QORG J011619.6$-$615344, $p_\mathrm{QSO} = 0.98$ &  & \\
 & $3.4^\prime$ & SUMSS J011656$-$615013: $43(2)\,\mathrm{mJy/beam}^\S$ & -- & -- & \\
 & \multirow{2}{*}{$3.7^\prime$} & \multirow{2}{*}{SUMSS J011643$-$615653: $33(1)\,\mathrm{mJy/beam}^\S$} & 0280-0013875: $20.0^\mathrm{m}/15.1^\mathrm{m}$ & \multirow{2}{*}{--} & \\
  &  &  & 0280-0013876: $18.0^\mathrm{m}/15.0^\mathrm{m}$ &   & \\
\hline
 J0143.6$-$5844 & $1.6^\prime$ & SUMSS J014347$-$584550: $27(1)\,\mathrm{mJy/beam}^\S$ & 0312-0011298: $18.5^\mathrm{m}/16.6^\mathrm{m}$ & SWIFT J014347.3$-$584551: 11.5(3)$^\mathdollar$ & \boldmath$\surd$ \\
\hline
\multirow{6}{*}{J0305.0$-$1602} &
 $2.3^\prime$ & NVSS J030511$-$160249: $7(1)\,\mathrm{mJy}$ & -- & -- & \\
 & \multirow{2}{*}{$3.2^\prime$} & \multirow{2}{*}{NVSS J030509$-$160450: $3(1)\,\mathrm{mJy}$} & 0739-0030568: $21.6^\mathrm{m}/18.9^\mathrm{m}$ & \multirow{2}{*}{--} & \\
  &  &  & 0739-0030569: $20.6^\mathrm{m}/18.8^\mathrm{m}$\,$^\dag$ &   & \\
 & $5.4^\prime$ & NVSS J030442$-$160458: $8(2)\,\mathrm{mJy}$ & -- & -- & \\
 & $5.8^\prime$ & NVSS J030521$-$160525: $47(2)\,\mathrm{mJy}$ & -- & -- & \\
 & \multirow{2}{*}{$6.9^\prime$} & \multirow{2}{*}{NVSS J030515$-$160814: $917(35)\,\mathrm{mJy}$} & \multirow{2}{*}{0738-0029664: $19.6^\mathrm{m}/17.0^\mathrm{m}$} & 2RXP J030515.3$-$160811: $1.26$ & \\
  &  &  &  & SWIFT J030514.9$-$160818: 1.5(1)$^\mathdollar$ & \\
\hline
\multirow{4}{*}{J0312.8+2013} &
 $2.4^\prime$ & NVSS J031247+201606: $5.1(4)\,\mathrm{mJy}$ & -- & -- & \multirow{4}{*}{\boldmath$\surd$ \cite{2012ApJ...752...61M}} \\
 & $2.8^\prime$ & NVSS J031240+201141: $18(1)\,\mathrm{mJy}$ & 1101-0034847: $21.2^\mathrm{m}/19.4^\mathrm{m}$ & -- & \\
 & $4.7^\prime$ & NVSS J031307+201229: $2.7(4)\,\mathrm{mJy}$ & -- & -- & \\
 & $5.6^\prime$ & NVSS J031245+201913: $4(1)\,\mathrm{mJy}$ & 1103-0035146: --$/19.4^\mathrm{m}$ & -- & \\
\hline
\multirow{2}{*}{J0338.2+1306} &
 $5.1^\prime$ & NVSS J033803+131045: $28(1)\,\mathrm{mJy}$ & -- & -- & \\
 & $5.5^\prime$ & NVSS J033829+130215: $15.1(6)\,\mathrm{mJy}$ & 1030-0045117: $19.3^\mathrm{m}/18.3^\mathrm{m}$ & SWIFT J033829.2+130217: 1.5(4)$^\mathdollar$ & \boldmath$\surd$\\
\hline
\multirow{2}{*}{J0438.0$-$7331} &
 $3.3^\prime$ & SUMSS J043836$-$732921: $20(1)\,\mathrm{mJy/beam}^\S$ & 0165-0071403: $15.6^\mathrm{m}/14.1^\mathrm{m}$ & -- & \\
 & $6.3^\prime$ & SUMSS J043900$-$732648: $63(2)\,\mathrm{mJy/beam}^\S$ & 0165-0071476: $17.8^\mathrm{m}/17.0^\mathrm{m}$ & -- & \\
\hline
 J0737.5$-$8246 & $2.3^\prime$ & SUMSS J073706$-$824836: $14(1)\,\mathrm{mJy/beam}^\S$ & 0071-0020954: $17.8^\mathrm{m}/17.5^\mathrm{m}$ & -- & \boldmath$\surd$ \cite{2012ApJ...752...61M}\\
\hline
 J1223.3+7954 & $1.9^\prime$ & NVSS J122358+795329: $31(1)\,\mathrm{mJy}$ & 1698-0045483: $20.2^\mathrm{m}/18.5^\mathrm{m}$ & -- & \\
\hline
\multirow{3}{*}{J1347.0$-$2956} &
 $2.2^\prime$ & NVSS J134706$-$295840: $27(1)\,\mathrm{mJy}$ & 0600-0304193: $18.8^\mathrm{m}/17.1^\mathrm{m}$ & 1WGA J1347.1$-$2958: $0.23(4)$ & \multirow{3}{*}{\boldmath$\surd$ \cite{2012ApJ...752...61M}}\\
 & $3.5^\prime$ & NVSS J134653$-$295346: $12(1)\,\mathrm{mJy}$ & 0601-0302620: $19.8^\mathrm{m}/18.4^\mathrm{m}$ & -- &  \\
 & $4.7^\prime$ & -- &  & 1WGA J1347.0$-$2952: $0.28(5)$ & \\
\hline
 J1410.4+7411 & -- & -- & -- & -- & \boldmath$\surd$ \\
\hline
\multirow{6}{*}{J2257.9$-$3646} & 
 $0.2^\prime$ & -- & 0532-0895676: $19.9^\mathrm{m}/18.4^\mathrm{m}$ & 1WGA J2257.9$-$3645: $0.08(1)$ & \\
 & $3.7^\prime$ & NVSS J225741$-$364833: $6.9(5)\,\mathrm{mJy}$ & 0531-0930900: $19.8^\mathrm{m}/18.9^\mathrm{m}$ & -- &  \\
 & $4.3^\prime$ & NVSS J225815$-$364433: $10.6(6)\,\mathrm{mJy}$ & 0532-0895730: $19.2^\mathrm{m}/18.1^\mathrm{m}$ & 1WGA J2258.2$-$3644: $0.21(2)$ & \\
 & $4.5^\prime$ & NVSS J225817$-$364520: $7.4(5)\,\mathrm{mJy}$ & 0532-0895735: $20.7^\mathrm{m}/18.5^\mathrm{m}$ & -- &  \\
 & $5.2^\prime$ & -- & multiple & 1WGA J2258.3$-$3647: $0.05(1)$ & \\
 & $7.0^\prime$ & -- & 0533-0862976: $19.7^\mathrm{m}/19.8^\mathrm{m}$ & 1WGA J2258.0$-$3638: $0.027(5)$ & \\
\hline
J2347.2+0707 & $3.9^\prime$ & NVSS J234706+070351: $40(2)\,\mathrm{mJy}$ & 0970-0695942: $21.2^\mathrm{m}/20.5^\mathrm{m}$ & -- &  \\
\hline
 \end{tabular}
 \caption{Counterpart candidates, sorted by increasing angular
   separation from the nominal 2FGL position. Radio and X-ray sources
   are located within the 95\% positional uncertainty of the 2FGL
   source. The radio flux is given in mJy at 1.4\,GHz \mbox{($^\S$: 843\,MHz)}, 
   while the unabsorbed X-ray flux between 0.2--2\,keV 
   ($^\mathdollar$: 0.3--2\,keV) is listed in 
   $10^{-12}\,\mathrm{erg}\,\mathrm{cm}^{-2}\,\mathrm{s}^{-1}$. Apart
   from the \emph{Swift} sources, the X-ray flux was derived from the
   cataloged count rates (2RXP \cite{2001ROSAT}, 1WGA
   \cite{2000yCat.9031....0W}), assuming a power-law with index
   $\Gamma=2.0$ and a hydrogen column density $N_H$ as obtained from
   the LAB survey, see section \ref{ssec:XRT_ana} (with WebPIMMS,
   http://heasarc.gsfc.nasa.gov/Tools/w3pimms.html). For every radio
   or X-ray source, the table lists corresponding optical counterpart
   candidates along with optical flux in USNO B2/R2 [$^\dag$: B1/R2,
     $^\star$: SDSS $r$] magnitudes (HB2012a
   \cite{2010AJ....140.1987H}, QORG
   \cite{2004A&A...427..387F}). Parentheses indicate the error on the
   last decimal. The last column indicates sources associated with
   infrared (WISE) blazar candidates, see section
   \ref{ssec:WISE_assoc} in this paper and
   \cite{2012ApJ...752...61M}.} \label{tab:mw_assoc}
\end{center}
\end{scriptsize}
\end{sidewaystable}

\newpage
\begin{sidewaystable}[H]
\begin{center}
\begin{scriptsize}
\begin{tabular}{|llcccc|}
\hline
 2FGL name & WISE name & $W1\,[\mathrm{mag}]$ & $W2\,[\mathrm{mag}]$ & $W3\,[\mathrm{mag}]$ & $W4\,[\mathrm{mag}]$ \\
\hline
J0031.0+0724 & J003119.70+072453.6 & 13.89(3) & 13.06(3) & 10.58(11) & 8.47(39) \\
J0143.6$-$5844 & J014347.39$-$584551.3 & 13.39(2) & 12.70(2) & 10.77(6) & 8.88(28) \\
J0305.0$-$1602 & J030515.07$-$160816.5 & 14.21(3) & 13.64(3) & 12.21(29) & 9.36 \\
J0338.2+1306 & J033829.26+130215.6 & 13.83(3) & 13.04(3) & 11.16(16) & 8.30 \\
J1410.4+7411 & J141235.75+741158.0 & 15.03(3) & 14.28(4) & 12.23(18) & 9.55 \\
\hline
\end{tabular}
\caption{Infrared sources detected with WISE
  \cite{2012yCat.2311....0C}, positionally coinciding with the radio
  and X-ray associations established for the preselected 2FGL
  candidate sample. We list the WISE magnitudes $W1, W2, W3,
  \textnormal{ and } W4$, corresponding to the $3.4, 4.6, 12,
  \textnormal{ and } 22\,\upmu\mathrm{m}$ bands. The error on the last
  decimals is written in parentheses. If no error is given, the value
  represents an upper limit (95\,\% confidence
  level).} \label{tab:WISE_assoc}
\end{scriptsize}
\end{center}
\vspace{0.75cm}
\begin{center}
\begin{scriptsize}
\begin{tabular}{|llcccccccc|}
\hline
\multirow{2}{*}{Obs.\,ID} & \multicolumn{1}{c}{Name} & $\sigma_{90}$ & \multirow{2}{*}{$S/N$} & $N_\mathrm{H}$ & $f^\mathrm{abs}(0.3\!-\!2\,\mathrm{keV})$ & $\phi_0$ & \multirow{2}{*}{$\Gamma$} & \multirow{2}{*}{$C_\mathrm{stat}/$dof} & $f^\mathrm{unabs}(0.3\!-\!2\,\mathrm{keV})$ \\ 
  & \multicolumn{1}{c}{SWIFT} & [arcsec] &  & [$10^{20}\,\mathrm{cm}^{-2}$] & [$10^{-13}\,\mathrm{erg}\,\mathrm{cm}^{-2}\,\mathrm{s}^{-1}$] & [$10^{-4}\,\mathrm{keV}^{-1}\,\mathrm{cm}^{-2}\,\mathrm{s}^{-1}$] &  &  & [$10^{-13}\,\mathrm{erg}\,\mathrm{cm}^{-2}\,\mathrm{s}^{-1}$] \\ 
\hline
\multirow{3}{*}{41274} &
J014229.0$\!-\!$584553 & $5$ & $7.4$ & $2.13$ & $3.03_{-0.48}^{+0.50}$ & $1.21_{-0.18}^{+0.20}$ & $1.56_{-0.22}^{+0.23}$ & $9.9/10$ &  $3.37_{-0.56}^{+0.61}$ \\ 
 & J014347.3$\!-\!$584551 & $4$ & $47.4$ & $2.04$ & $100.02_{-2.53}^{+2.58}$ & $35.76_{-0.80}^{+0.81}$ & $2.20_{-0.03}^{+0.03}$ & $125.3/145$ & $115.10_{-3.03}^{+3.06}$ \\ 
 & J014410.1$\!-\!$584042 & $6$ & $4.1$ & $2.03$ & $0.86_{-0.22}^{+0.22}$ & $0.32_{-0.08}^{+0.09}$ & $2.00$ & $2.2/3$ & $0.97_{-0.24}^{+0.28}$ \\
\multirow{1}{*}{41286} &
J030514.9$\!-\!$160818 & $4$ & $13.1$ & $3.62$ & $12.18_{-1.06}^{+1.14}$ & $5.03_{-0.40}^{+0.42}$ & $1.92_{-0.12}^{+0.12}$ & $26.9/30$ &  $14.98_{-1.43}^{+1.48}$ \\
\multirow{2}{*}{41292} &
J033829.2$+$130217 & $5$ & $6.5$ & $15.30$ & $5.65_{-0.87}^{+1.02}$ & $3.68_{-0.54}^{+0.59}$ & $2.73_{-0.32}^{+0.31}$ & $6.1/7$ &  $14.58_{-3.39}^{+3.51}$ \\
 & J033840.4$+$130722 & $5$ & $4.4$ & $15.20$ & $1.00_{-0.21}^{+0.22}$ & $0.65_{-0.13}^{+0.15}$ & $2.00$ & $3.2/4$ &  $1.99_{-0.40}^{+0.46}$ \\
\multirow{1}{*}{47219} &
 J141234.8$+$741153 & $5$ & $5.4$ & $2.35$ & $2.33_{-0.46}^{+0.55}$ & $0.85_{-0.15}^{+0.16}$ & $2.21_{-0.28}^{+0.29}$ & $11.1/5$ &  $2.74_{-0.58}^{+0.63}$ \\
\hline
\end{tabular}
\caption{List of X-ray sources detected with \emph{Swift}-XRT, sorted
  by right ascension. Apart from positional information (SWIFT
  JHHMMSS.s$\pm$DDMMSS, where $\sigma_{90}$ denotes the uncertainty at
  $90\%$ confidence), the table lists the signal-to-noise ratio of the
  absorbed flux $f^\mathrm{abs}$ between 0.3 and 2\,keV for each
  observation (Obs.\,ID). The spectra were fit with a power-law model
  corrected for photoelectric absorption (normalization $\phi_0$,
  power-law index $\Gamma$), fixing the hydrogen column density
  $N_\mathrm{H}$ to the nominal Galactic value. The C-statistic as
  implemented in \emph{Xspec} was used for spectral fitting
  ($C_\mathrm{stat}$, dof denotes the number of degrees of
  freedom). The unabsorbed flux $f^\mathrm{unabs}$ between 0.3 and
  2\,keV was derived from the power-law fit.} \label{tab:swift_assoc}
\end{scriptsize}
\end{center}
\end{sidewaystable}

\end{appendix}

\bibliographystyle{JHEP}
\bibliography{ms.bbl}

\end{document}